\documentclass[prb,aps,twocolumn]{revtex4}
\usepackage{amsmath}
\usepackage{amssymb}
\usepackage {graphicx}
\usepackage{color}
\usepackage{bbold}

\newcommand{\beq}{\begin{equation}}
\newcommand{\eeq}{\end{equation}}
\newcommand{\bea}{\begin{eqnarray}}
\newcommand{\eea}{\end{eqnarray}}
\newcommand{\e}{\varepsilon}
\newcommand{\bk}{{\vec k}}

\newcommand{\bq}{{\vec q}}

\begin{document}

\title{Conservation laws, vertex corrections, and screening in Raman spectroscopy}
\author{Saurabh Maiti$^1$, Andrey V. Chubukov$^2$ and P. J. Hirschfeld$^1$}
\affiliation{$^1$Department of Physics, University of Florida, Gainesville, FL 32611}
\affiliation{$^2$Department of Physics, University of Minnesota, Minneapolis, Minnesota 55455, USA}

\date{\today}
\begin{abstract}
We present a microscopic theory for the Raman response of a clean
multiband superconductor, with emphasis on the effects of vertex
corrections and long-range Coulomb interaction. The measured Raman
intensity, $R(\Omega)$, is proportional to the imaginary part of
the fully renormalized particle-hole correlator  with Raman
form-factors $\gamma(\bk)$. In a BCS superconductor, a bare Raman
bubble is non-zero for any $\gamma(\bk)$ and diverges at $\Omega =
2\Delta +0$, where $\Delta$ is the largest gap along the Fermi
surface. However, for $\gamma(\bk) =$ const, the full $R(\Omega)$
is expected to vanish due to particle number conservation. It was
long thought that this vanishing is due to the singular screening
by long-range Coulomb interaction.  We show diagrammatically that
this vanishing actually holds due to vertex corrections from the
same short-range interaction that gives rise to superconductivity.
We further argue that long-range Coulomb interaction does not
affect the Raman signal for {\it any}  $\gamma(\bk)$. We argue
that vertex corrections eliminate the divergence at $2\Delta$ and
replace it with a maximum at a somewhat larger frequency.  We also
argue that vertex corrections give rise to sharp peaks in
$R(\Omega)$ at $\Omega < 2\Delta$, when $\Omega$ coincides with
the frequency of one of collective modes in a superconductor, e.g,
Leggett and Bardasis-Schrieffer modes in the particle-particle
channel , and an excitonic mode in the particle-hole channel.
\end{abstract}
\maketitle
\section{Introduction}\label{sec:intro}
Raman spectroscopy is a useful tool to probe the electronic
properties of a correlated  metal. It is specifically important
for superconductors (SC) because it can probe not only
particle-hole excitations, but also particle-particle fluctuations
of the condensate, making it an extremely valuable probe to study
collective excitations in a SC, both in the dominant and in the
subdominant pairing channel. Raman scattering  can probe
fluctuations in any scattering geometry,  regardless of the
symmetry of the SC state. This unique ability of Raman
spectroscopy is due to the fact that a given geometry can be
selected by choosing the polarizations of the incoming and the
scattered light relative to the crystallographic
axes.\cite{Sastry1990,Deveaux}

If the energy of the incident and scattered light is smaller than
the gaps between the bands which cross the Fermi level and those
which don't, the would be resonant scattering between these two
types of bands is absent, and the Raman intensity, $R(\Omega)$,
can be evaluated in the non-resonant approximation, where it is
proportional to the imaginary part of the fully renormalized
correlation function of modulated densities of fermions from the
bands which cross the Fermi level: $R(\Omega) \propto {\rm Im}
\left[\sum_{\bk}\langle\rho^R(\bk,0)\rho^R(\bk,\Omega)
\rangle\right]$, where $\rho^R(\bk)=\sum_a\rho^R_a(\bk)$, $a$
labels the bands, and $\rho_R^{a} (\bk) = \gamma^a (\bk)c^\dag_a
(\bk) c_a (\bk)$. The form-factor  $\gamma^a(\bk)$ (also called
the Raman vertex) is expressed in terms of particular components
of the effective mass tensor\cite{Deveaux} which depends on the
polarization of the incoming and outgoing light. In the absence of
a sizable overlap  between density operators  with the same $\bk$
from different bands, $R(\Omega)$ can be further approximated by
$\sum_{\bk,a} \langle\rho_R^{a} (\bk,0) \rho_R^{a} (\bk, \Omega)
\rangle$.

Non-resonant Raman spectroscopy has been used extensively to
extract the pairing gap $\Delta$ and analyze various collective
modes in the particle-particle channel, like the
Leggett\cite{Leggett} and Bardasis Schrieffer(BS) modes.
\cite{BardasisSchrieffer61,vaks,zawadowskii} The former
corresponds to  fluctuations of the relative phases of the
condensed order parameters in a multiband superconductor, and the
latter corresponds to gapped fluctuations of an un-condensed order
parameter in a subleading attractive pairing channel (e.g.
fluctuations of a $d-$wave order parameter in an $s-$wave SC if
both $s-$ wave and $d-$wave channels are attractive, but
attraction in the $d-$wave is weaker than that in the s-wave
channel). The Leggett mode has been reported to be observed in
Raman experiments on MgB$_2$ ~\cite{MGB2,kkklein} in the A$_{\rm
1g}$ scattering geometry.  BS modes have not been observed in
conventional SCs, presumably because the competing attractive non
$s$-wave pairing channels are too weak.  However, a BS mode was
predicted in  Fe-based SC, because in many of these materials the
$d-$wave channel is attractive, and the attraction is sometimes as
strong as the one in the $s-$wave channel.\cite{rev1,rev2} Raman
experiments on hole doped Ba$_{1-x}$K$_x$Fe$_2$As$_2$
\cite{Hackl12,Hackl14} and electron-doped
NaFe$_{1-x}$Co$_x$As\cite{Blumberg14} have reported features in
the d-wave (B$_{\rm 1g}$)  channel, consistent with a BS mode.

There are also two other collective modes in a superconductor. One
is a Boguliubov-Anderson-Goldstone (BAG) mode of phase
fluctuations, associated with spontaneously broken U(1) symmetry
(fluctuations  of the overall phase of different condensates in
case of a multiband SC).  The BAG mode is massless in a
charge-neutral superfluid, but becomes a plasmon in a charged
superconductor due to long-range Coulomb interaction.  The other
is an amplitude  mode of a condensate\cite{L0,L1,L2}, often called
Higgs mode by analogy with the massive boson mode in the Standard
Model of particle physics.  The  amplitude and the phase modes
decouple in a BCS superconductor in the absence of time reversal
symmetry breaking.\cite{spisLegett2,spisLegget1,SM_AVC,Benfatto}
Neither of these modes is, however, Raman-active, unless special
conditions are met. The phase mode only contributes to the Raman
intensity with a weight $\propto q^2$, where $q$ is the momentum
at which the Raman signal is measured. In Raman experiments, the
momentum $q$ is smaller by $v_F/c$ than a typical fermionic
momentum $k \sim \Omega/v_F$. Accordingly, the spectral weight of
the phase mode contribution to the Raman intensity is very small.
The amplitude does not directly couple to $\rho_R$, and appears in
the Raman response only when there is an interaction with other
collective modes like phonons  or magnons,\cite{Zwerger} and/or
when superconductivity  emerges out of a pre-existing
charge-density-wave state.\cite{L0,Little} In this work we do not
take phonons or magnons into consideration, and do not assume
pre-existing density-wave order.

The theory of non-resonant Raman response in superconductors has a
long
history.\cite{ABGenkin,Klein,ABF,zawadowskii,griffin,devereaux}
For a superconductor with a minimal gap $\Delta$, the Raman
intensity $R(\Omega)$, computed within BCS theory, as the
imaginary part of a bare particle-hole bubble with $\gamma^a
(\bk)$ in the vertices, is non-zero at  $\Omega > 2\Delta$ and has
an edge singularity at $\Omega = 2\Delta +0$  in all scattering
geometries. This holds  even if $\gamma^a (\bk)$ is a constant.
For a nodal superconductor, $R(\Omega)$ is non-zero for all
frequencies and has a singularity at $\Omega = 2\Delta_{\rm max}$,
where $\Delta_{\rm max}$ is the maximum gap.  In this respect, the
behavior of a bare particle-hole bubble in a superconductor
differs from that in the normal state, where the free-fermion
particle-hole bubble vanishes in the limit when $\Omega$ is finite
and $q \to 0$ (more specifically, when $\Omega \gg v_F q$). This
vanishing in the normal state is related to particle number
conservation and holds for any $\gamma^a(\bk)$ because for free
fermions each $n_k$ in $N = \sum_{\bk} n_{\bk}$ is separately
conserved.  A non-zero value of this bubble in a superconductor is
the consequence of the fact that a BCS Hamiltonian formally does
not conserve the number of fermions due to the presence of
$c^\dagger_k c^\dagger_{-k}$ and $c_k c_{-k}$ terms.

\begin{figure}[htp]
\includegraphics[width=0.9\columnwidth]{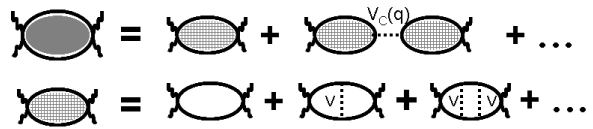}
\caption{ \label{fig:1} The summation scheme  to calculate the
non-resonant Raman response. The top line is the RPA sum involving
the long-range Coulomb interaction $V_C(\bq)$. The bottom line
denotes the ladder approximation to account for the vertex
corrections to the bare bubble. Here $V$ denotes a generic
short-range interaction.}
\end{figure}

Several groups argued~\cite{dev_2,Deveaux,Girsh,PeterBoyd} that
$R(\Omega)$ in one-band superconductor indeed vanishes for $\gamma
(\bk) =$ const, once one adds to BCS Hamiltonian the 4-fermion
interaction term  describing long-range Coulomb interaction $V_C
(\bq)$. This interaction renormalizes $R(\Omega)$  by adding
RPA-type series of particle-hole bubbles  coupled by $V_C (\bq)$
[the upper line in Fig. \ref{fig:1}]. In a two-band
superconductor, a similar consideration \cite{Girsh} yielded a
partial reduction of $R(\Omega)$ when $\gamma (\bk)$ are
momentum-independent, but not equal for the two bands.

This point of view has been challenged recently by Cea and
Benfatto~\cite{Lara}. They  used  gauge-invariant effective action
approach and computed the Raman response of a one-band and
two-band  $s-$wave SC in A$_{\rm 1g}$  geometry. They argued that
the total number of fermions, including fermions in the
condensate, is a conserved quantity.  As the result, when
$\gamma^a (\bk)$ is a constant, independent of $a$, the fully
dressed $R(\Omega)$ must vanish already before one includes the
renormalizations due to Coulomb interaction, because in this case
Raman susceptibility coincides with the density susceptibility,
and the latter must vanish at $q=0$ and finite $\Omega$ due to
conservation of the total number of particles (or, equivalently,
of the total charge).

In this communication we analyze Raman response of one-band and
multi-band  superconductors with various pairing symmetry  using a
direct diagrammatic approach. We argue that the full
gauge-invariant diagrammatic analysis of Raman intensity in a
superconductor necessarily includes the processes which
renormalize a given  particle-hole bubble (the lower line in Fig.
\ref{fig:1}). We show, in agreement with Ref. [\onlinecite{Lara}],
that these renormalizations give rise to the vanishing of
$R(\Omega)$ for $\gamma (\bk) =$ const even before one adds
long-range Coulomb interaction. Moreover, we argue that long-range
Coulomb interaction is completely irrelevant for Raman scattering
at vanishing $q$ and finite $\Omega$, because RPA renormalizations
between dressed bubbles only give contributions to $R(\Omega)$
which scale as  at least as $q^2$.

\begin{figure}[htp]
\includegraphics[width=0.7\columnwidth]{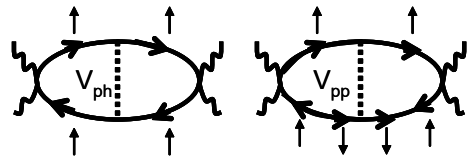}
\caption{ \label{fig:2} Diagrams in the ladder series for the
Raman response that we call in the text particle-hole and
particle-particle type contributions.}
\end{figure}

We treat a superconductor within a weak coupling approach. In this
limit, the fermionic self-energy is irrelevant, and essential
renormalizations within a particle-hole bubble come from the
ladder series of vertex corrections. We categorize these  vertex
corrections into two categories -- particle-hole and
particle-particle contributions. The first ones involve pairs of
fermionic Green's functions with opposite direction of arrows, the
second ones involve pairs of fermionic Green's functions with the
same direction of arrows, as shown in Fig. \ref{fig:2}. These
combinations appear in a superconductor once a particle-hole pair,
which couples to the light, gets converted into a
particle-particle pair via a process which propagates as one
normal and one anomalous Green's function. We show that the
conservation of the total number of particles guarantees certain
cancellations between the bare bubble and the renormalizations
from vertex corrections in the particle-particle channel. As a
result:

\begin{itemize}
\item For a constant $\gamma^a ({\bk})$, $R(\Omega)$ for a one-band
superconductor vanishes once one includes  vertex corrections in
the particle-particle channel. This holds even if we additionally
include vertex corrections in the particle-hole channel (see also
Ref.  [\onlinecite{Lara}]).
\item For a momentum-dependent $\gamma^a ({\bk})$, $R(\Omega)$ is
non-zero, but the ``$2\Delta$'' edge singularity is removed and
replaced by a maximum at an energy \emph{above} $2\Delta$.  This
holds for isotropic or anisotropic systems and all scattering
geometries.
\item In a multi-band superconductor, $R(\Omega)$ vanishes due to vertex
corrections in the particle-particle channel, when  $\gamma^a
({\bk})$ has the same constant value for all bands (see also Ref.
[\onlinecite{Lara}]).
\item When  $\gamma^a ({\bk})$ is either momentum-dependent, or has
different constant values for different bands, $R(\Omega)$ is
non-zero, but the ``$2\Delta$'' edge singularity is again
eliminated and replaced by a maximum at a frequency above
$2\Delta$.
\item In both one-band and multi-band nodeless superconductors,
$R(\Omega)$ generally vanishes below twice the minimum gap
$2\Delta^a$, but under proper conditions may have
$\delta-$function contributions at $\Omega < 2\Delta^a$ from
Leggett and BS-type modes in the particle-particle channel, and
from excitonic modes in the particle-hole channel. In nodal
superconductors, $R(\Omega)$ is finite at all frequencies and the
contributions from collective modes appear as resonance peaks with
a finite width.
\item The
Coulomb interaction does not affect $R(\Omega)$ at $\Omega \gg v_F
q\rightarrow 0$. For non s-wave scattering geometry, this holds
due to symmetry reasons and is well-known.  We argue that this
also holds for $s-$wave scattering geometry in one- and multi-band
systems, and in isotropic and lattice systems (where the $s-$wave
Raman vertex $\gamma^a_{\bk}$ is momentum-dependent and the Raman
bubble is different from a density-density bubble).
\end{itemize}
Vertex corrections inside a particle-hole bubble have been
analyzed in the past. The Raman response in a 1-band SC was worked
out in Ref. [\onlinecite{Klein}]. Our formulas fully agree with
the ones in their work, although we interpret the results
differently.

A bilayer superconductor was analyzed in Ref. [\onlinecite{dev_2}]
and a 2-band SC with Fermi surfaces separated in momentum space
was analyzed in Ref. [\onlinecite{chubilya}]. The authors of Ref.
[\onlinecite{chubilya}]  obtained the full expression for the
Raman intensity $R(\Omega)$ with vertex corrections in the
particle-hole and particle-particle channel and also included
renormalizations due to Coulomb interaction. However, they
analyzed the results only  for a particular momentum-dependent
$\gamma^a ({\bk})$ and for small frequencies $\Omega \ll 2\Delta$
and didn't address the behavior of $R(\Omega)$ near $2\Delta$. The
authors of Ref. [\onlinecite{Lara}] considered both one-band and
two-band superconductors and specifically singled out the
contribution to the Raman intensity from the Leggett mode.  The
authors of  Ref. [\onlinecite{DevereauxScalapino09}] considered
the effects of vertex corrections in the B$_{1g}$ channel and put
the  emphasis on the contribution of the BS mode.  The authors of
Ref. [\onlinecite{KhodasChubukov}] analyzed vertex corrections in
the particle-hole channel and studied the resulting excitonic
modes, and the authors of Ref. [\onlinecite{SMPJH}] discussed the
possibility of multiple B$_{1g}$ BS modes.

The goals of our work are three-fold. First, to show how to carry
out a gauge-invariant calculation of the Raman response
diagrammatically, starting from a microscopic model. We argue that
vertex corrections in the particle-particle channel must be
included for this purpose; Second, to analyze the interplay
between vertex corrections in the particle-particle and
particle-hole channels.  Both types of corrections lead to
collective modes, and we show that the interplay between them is
rather complex; Third, to analyze the effect of long-range Coulomb
interaction on the Raman bubble, already dressed by vertex
corrections.

We obtain the generic expression for the Raman intensity,  valid
for any number of bands, any scattering geometry, and any
spin-singlet gap symmetry. For demonstration purposes, later in
the paper we focus on one-band and two-band 2D $s-$wave
superconductors on a square lattice, in A$_{\rm 1g}$ scattering
geometry.

The general result, when applied to the one-band model, reproduces
the result of Ref. [\onlinecite{Lara}] that the Raman intensity
$R(\Omega)$ vanishes for a constant Raman vertex ($\gamma(\bk) =$
const) due to vertex corrections. As an extension to that work, we
expand the pairing interaction and the Raman vertex in $A_{1g}$
harmonics. We show that $R(\Omega)$ is finite when $\gamma (\bk)$
has momentum dependence (e.g., $\cos {4\theta}$ harmonic in a 2D
system on a square lattice). We show that the $2\Delta$ edge
singularity in this $R(\Omega)$  is eliminated by vertex
corrections, as long as the pairing interaction is also
momentum-dependent, and is replaced by a maximum at a frequency
above $2\Delta$. We argue that $R(\Omega)$ may also have
$\delta$-function peaks at $\Omega < 2\Delta$ due to BS-type and
excitonic collective modes.

When applied to the two-band model, we reproduce another result of
Ref. [\onlinecite{Lara}]  that $R(\Omega)$  is non-zero when the
Raman vertices $\gamma^a (\bk)$, $a=1,2$, are different constants
for the two bands. In addition to that work we also consider the
case when $\gamma^a (\bk)$ are momentum-dependent. We show that
the $2\Delta$ peak in $R(\Omega)$  is again eliminated by vertex
corrections and replaced by a maximum at a higher frequency. We
analyze potential $\delta$-function peaks at $\Omega < 2\Delta$
due to collective modes: Leggett collective mode in the
particle-particle channel and excitonic collective mode in the
particle-hole channel. In this analysis, we reproduce and
generalize the results of Refs. [\onlinecite{Lara}] and
[\onlinecite{chubilya}], respectively. We additionally consider
the effects due to BS-type collective modes. We analyze in detail
the interplay between the effects from collective modes in the
particle-particle and particle-hole channels.

We next analyze the effect of the long-range Coulomb interaction.
In graphical representation, this interaction creates series of
additional renormalizations of the Raman bubble. All terms in
these series contain the square of a fully renormalized bubble
with the Raman vertex on one side and a total density vertex on
the other. We demonstrate that such a bubble vanishes and hence
the Coulomb interaction does not contribute to Raman scattering,
even in $A_{1g}$ scattering geometry. We demonstrate this
explicitly for the one-band model for a general $\gamma(\bk)$, and
for the two-band model for the case when $\gamma^a (\bk)$ is
momentum independent, but has different values for the two bands.

We also discuss some specific examples of the Raman scattering in
2D square-lattice systems in non-A$_{\rm 1g}$ scattering geometry.
In particular, we argue that for a $d-$wave superconductor, the
Raman vertex in A$_{\rm 1g}$ scattering geometry with $\gamma
(\bk) =$ const vanishes due to vertex renormalizations which
involve the $d-$wave component of the interaction in the
particle-particle channel, i.e., the one that gives rise to the
pairing; while in a B$_{\rm 1g}$ scattering geometry  (with, e.g.,
$\gamma (\bk) \propto \cos{2\theta}$), the renormalization of the
bare Raman bubble in the particle-particle channel involves the
$s$-wave component of the interaction in the particle-particle
channel.

The rest of the text is organized as follows. In Sec
\ref{sec:interact}, we introduce an effective low energy model. In
Sec \ref{sec:Raman}, we describe the generic computational scheme
to calculate diagrammatically the Raman response with vertex
corrections and screening. In Sec. \ref{sec:ex}, we  apply this
scheme to analyze vertex corrections within the Raman bubble. We
consider one-band and two-band  $s$-wave superconductors and
$A_{1g}$ Raman scattering geometry and analyze the condition under
which $R(\Omega)$ vanishes, the elimination of ``$2\Delta$"  edge
singularity, and the effects due to Raman-active collective modes.
In Sec. \ref{sec:screening}, we discuss the role of Coulomb
interaction and argue that it does not affect Raman scattering. In
Sec. \ref{sec:learn}, we briefly discuss Raman intensity in other
scattering geometries and for other gap symmetries, and consider a
specific example of a $d-$wave superconductor and $A_{1g}$ and
$B_{1g}$ scattering geometries. We present our conclusions in Sec.
\ref{sec:conclusions}.

\begin{figure}[htp]
\includegraphics[width=0.9\columnwidth]{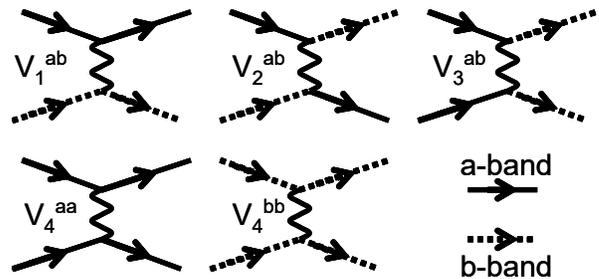}
\caption{ \label{fig:0} All possible interactions between a pair
of bands $a,b$ (solid and dashed respectively). $V^{ab}_1$ is the
density-density interaction between band $a$ and band $b$;
$V^{ab}_2$ is an exchange between the two bands; $V_3$ is a
Fermion-pair hopping term that is made possible due to Umpklapp
processes; $V^{ab}_{4,5}$ are the density-density interaction
terms within each band. Every pair of bands has a $V_{1,2,3}$
interaction component.}
\end{figure}

\section{The low energy model}\label{sec:interact}
We consider an effective low energy  model of a superconductor
with $N_b$ bands. The Hamiltonian is given by
$\mathcal{H}=\mathcal{H}_{0} + \mathcal{H}_{\rm int}$, where
$\mathcal{H}_{0}$ is the quadratic part, which includes the
superconducting condensate,  and $\mathcal{H}_{\rm int}$ is the
combination of 4-fermion interaction terms. We assume that the
pairing is in a spin-singlet channel, and that the condensates are
made out of pairs of fermions with momenta ${\bk}$ and $-{\bk}$
from the same band. The quadratic Hamiltonian $\mathcal{H}_{0}$ is
then diagonal in band basis: $\mathcal{H}_0 = \sum_{a=1}^{N_b}
\mathcal{H}_a$.  Throughout the paper we use indices $a$ and $b$
to label the bands. We use the Nambu formalism and  combine the
fermionic creation and annihilation operators into a Nambu spinor
$\Psi^{\dag}_a(\bk)\equiv
(c^{\dag}_{a,\bk,\uparrow},c_{a,-\bk,\downarrow},c^{\dag}_{a,\bk,\downarrow},
c_{a,-\bk,\uparrow})$. In this representation,
$\mathcal{H}_a=\sum_{\bk}\Psi^{\dag}_a (\bk) \mathcal{E}_a
(\bk)\Psi_a (\bk)$, where the $4\times4$ matrix
$\mathcal{E}_a(\bk)$ is given by \beq\label{eq:int_3}
\mathcal{E}_a(\bk)=\left(\begin{array}{cccc}
\e^{a}_{\uparrow}(\bk)&-\Delta^{a}_{\downarrow\uparrow}(\bk)&0&0\\
-\Delta^{a*}_{\downarrow\uparrow}(\bk)&-\e^{a}_{\downarrow}(-\bk)&0&0\\
0&0&\e^{a}_{\downarrow}(\bk)&-\Delta^{a}_{\uparrow\downarrow}(\bk)\\
0&0&-\Delta^{a*}_{\downarrow\uparrow}(\bk)&-\e^{a}_{\uparrow}(-\bk)
\end{array}\right)
\eeq The matrix $\mathcal{E}_a (\bk)$ can be written in more
compact form as $ \mathcal{E}_a (\bk)=\e^{a} Q^{a}_3 -
\Delta^{a}_R Q^{a}_1+ \Delta^{a}_I Q^{a}_2 $, where $Q^{a}_{i}=
\sigma_0 \otimes \sigma_{i}$ ($\sigma_0$ is $2\times2$ identity
matrix and $\sigma_i$ are the Pauli matrices), and
$\Delta_{R,I}^a$ are the real and imaginary parts of the pairing
gap $\Delta^{a}_{\downarrow\uparrow}(\bk)$. The spinor space is
spin$\otimes$Nambu (we use the same definition of the direct
product  as in Ref.~\onlinecite{direct}). The $4\times4$ Green's
function is then $G^a (\omega, \bk) =\left[i\omega
\sigma_0-\mathcal{E}_a({\bk})\right]^{-1}$. For spin-singlet
pairing, the $4\times4$ Nambu structure for $\mathcal{E}_a({\bk})$
reduces to two equivalent $2\times2$ structures, which differ by a
spin-flip.  We focus on one $2\times2$ structure [the upper left
$2\times2$ part in Eq. (\ref{eq:int_3})] and  drop the $\sigma_0$
spin component from the spinor space. Then  $\mathcal{E}_a$ and
$G$ become $2\times2$ matrices.  All formulas below will be
presented in this reduced space.

We approximate the interactions between low-energy band fermions
in a superconductor as functions of the momentum transfer $q$. In
doing so, we neglect the orbital composition of low-energy states,
i.e., the fact that in many cases (e.g.,
Fe-pnictides/chalcogenides)  band operators for low-energy states
are linear combinations of fermions from different
orbitals.\cite{rev2} Orbital physics generally induces additional
momentum dependence of the interaction potentials via form-factors
associated with the transformation from orbital to band basis for
each fermion involved in the interaction. This complication,
however, modifies the Raman response in a quantitative, but not in
a qualitative way.\cite{hcs}

We will need both unscreened and screened interactions  for the
computation of the Raman intensity. For the RPA renormalization of
the Raman bubble we need the interaction at the Raman momentum
transfer $q$. To avoid double counting we must treat this
interaction $V_C (q)$ as unscreened [e.g., $V_C(q)= 2\pi e^2/q$ in
2D].  For the renormalizations inside the Raman bubble (ladder
series of renormalizations) we need interactions with a momentum
transfer either of order $k_F$ or comparable to the distance $Q$
between different bands in momentum space, and  with energy
transfer comparable to $\Delta$. Such interactions should be taken
as the screened ones. We assume that $v_F Q > v_F k_F  > \Delta$,
in which case  screening  transforms a bare long-range static
Coulomb interaction into a short-range, but still static
interaction. Accordingly,  we approximate interactions inside the
Raman bubble by constants, different for $q \sim k_F$ (intra-band)
and $q \sim Q$ (inter-band).

In general, there are 4 types of intra-band and inter-band
short-range interactions between low-energy fermions:
density-density interaction between fermions from the same band
and density-density, exchange and pair-hopping interactions
between fermions from different bands (see Fig. \ref{fig:0}). We
use the same notations as in earlier
works\cite{ChubukovPhysica,AVCIE,RGSM} and label these
interactions as $V_4, V_1$, $V_2$, and $V_3$, respectively.
Interactions of each type have additional band indices as the ones
involving fermions from different bands are not necessarily equal.
The intra-band interaction $V_4$ is diagonal in band basis and we
label its components as $V^{aa}_4$. Interactions $V_1, V_2$, and
$V_3$ involve fermions from different bands, and we label their
components as $V^{ab}_1, V^{ab}_2$, and $V^{ab}_3$. There  are
${N_B}\choose{2}$ pairs of bands $(a,b)$ with $a\neq b$. In Nambu
notations the four interactions are
\begin{widetext}
\bea
\mathcal{H}_{\text{int}}&=& \frac{1}{2} \left(H_4 + H_1 + H_2 + H_3\right),\\
H_4 &=& \sum_{a} \sum_{\bq}  V^{aa}_4(q) \sum_{\bk}
\Psi^{\dag}_a(\bk) \sigma_3  \Psi_a(\bk+\bq)  \sum_{\bk'}
\Psi^{\dag}_a(\bk'+\bq) \sigma_3  \Psi_a(\bk') \label{eq:4} \\
H_1&=& \sum_{a \neq b} \sum_{\bq}  V^{ab}_1(q) \sum_{\bk}
\Psi^{\dag}_a(\bk) \sigma_3  \Psi_a(\bk+\bq)
\sum_{\bk'}\Psi^{\dag}_b(\bk'+\bq) \sigma_3  \Psi_b(\bk') \label{eq:4_1} \\
H_2 &=&\sum_{a\neq b}\sum_{\bq} V^{ab}_2(q) \times \nonumber \\
&&\sum_{\bk}
\left(\Psi^{\dag}_a(\bk) \sigma_+ \Psi_b(\bk+\bq) + \Psi^{\dag}_b(\bk) \sigma_-\Psi_a(\bk+\bq)\right)
\sum_{\bk'}\left(\Psi^{\dag}_b(\bk'+\bq) \sigma_+\Psi_a(\bk') + \Psi^{\dag}_a(\bk'+\bq) \sigma_-\Psi_b(\bk')\right) \label{eq:4_2} \\
H_3 &=&\sum_{a\neq b}\sum_{\bq} V^{ab}_3(q) \times \nonumber \\
&&\sum_{\bk}
\left(\Psi^{\dag}_a(\bk) \sigma_+ \Psi_b(\bk+\bq) + \Psi^{\dag}_b(\bk) \sigma_-\Psi_a(\bk+\bq)\right)
\sum_{\bk'}\left(\Psi^{\dag}_a(\bk'+\bq) \sigma_+ \Psi_b(\bk') + \Psi^{\dag}_b(\bk'+\bq) \sigma_- \Psi_a(\bk')\right)\label{eq:4_3}
\eea
where $\sigma_{\pm} = (\sigma_3 \pm \sigma_0)/2$. The long-range interaction with the bare Coulomb potential $V_{C} (q)$ is expressed as:
\bea\label{eq:Coulomb}
H_C&=& \frac{1}{4}  \sum_{\bq}  V_C (\bq) \sum_{a,b} \sum_{\bk}
\Psi^{\dag}_a(\bk) \sigma_3  \Psi_a(\bk+\bq)
\sum_{\bk'}\Psi^{\dag}_b(\bk'+\bq) \sigma_3  \Psi_b(\bk').
\eea
\end{widetext}
We also need self-consistency conditions on the pairing gaps
$\Delta^{a}_{\downarrow\uparrow} (\bk)$.  They represent the set
of coupled non-linear equations, which involve interactions
$V^{aa}_4$ and $V^{ab}_3$. In explicit form \bea\label{eq:self}
\Delta^a_{\downarrow\uparrow}(\bk)&=&-\int_{K'}V_4^{aa}(\bk-\bk')G^a_{12}(K')\nonumber\\
&&-\sum_{b\neq a}\int_{K'}V_3^{ab}(\bk-\bk')G^b_{12}(K'). \eea
where $K' = (\omega', \bk')$,  $\int_{K'}\equiv
T\sum_{\omega'}\int d^dk'/(2\pi)^d$, and $G^a_{12}$ is a
non-diagonal component of the $2\times 2$ matrix Green's function
$G^a (\omega, \bk) =\left[i\omega
\sigma_0-\mathcal{E}_a({\bk})\right]^{-1}$. We now proceed to
compute the Raman response.

\section{The Raman Response}\label{sec:Raman}
To calculate the fully renormalized Raman intensity we use the
computational scheme outlined in Fig. \ref{fig:1}. The Raman
intensity $R(\Omega)$ is proportional to the imaginary part of the
Raman susceptibility $\chi_R(\Omega)$. The latter is given by the
fully renormalized particle-hole bubble with Raman vertices on
both sides.  We use the short-range interaction from
$\mathcal{H}_{\rm int}$ for the renormalizations within a given
particle-hole bubble, and the long range Coulomb interaction  for
RPA renormalizations of the interaction-dressed bubbles (shaded
ones in Fig. \ref{fig:1}).  We approximate vertex renormalizations
within the bubble by the ladder series of vertex corrections.  We
argue that this procedure preserves gauge invariance, provided
that the equation for the SC gap is also obtained within the
ladder approximation. This scheme is a multiband generalization of
the computational approach used in Refs.
\onlinecite{Klein,zawadowskii}.

\begin{figure}[htp]
\includegraphics[width=0.9\columnwidth]{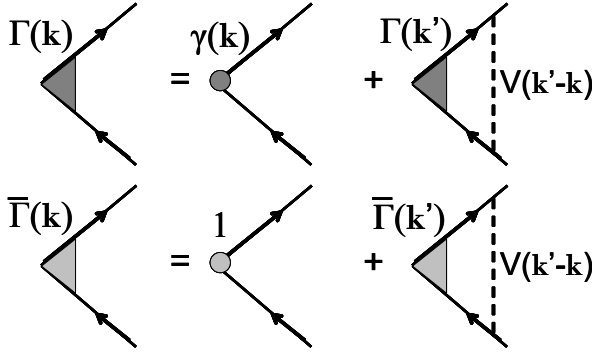}
\caption{\label{fig:01} Schematic form of the integral vertex equations for $\Gamma(\bk)$ and $\bar\Gamma(\bk)$ (band indices have been dropped). The `1' in the equation for $\bar\Gamma$ denotes the total density vertex (both $\gamma(\bk)$ and `1' are proportional to $\sigma_3$). Here $V(\bk'-\bk)$ represents all the relevant short range interactions that enter the renormalization of the respective vertices. The solid lines with arrows denote Green's function in Nambu space. The Green's functions in the Nambu space are matrices constructed of normal and anomalous Green's functions.}
\end{figure}

In the Nambu formalism, the Raman vertex for band $a$ is
$\gamma^a_{\bk}\sigma_3$ and the density vertex, which we will
need in RPA series, is $\sigma_3$.  The bare Raman susceptibility
is graphically represented by the particle-hole bubble with
$\gamma^a_{\bk}\sigma_3$ in the vertices: \beq \chi_{R,0} (Q) = -
\sum_{a}\int_K \text{Tr}\left[\gamma^a\sigma_3 {G}^a_K \gamma^a
\sigma_3 {G}^a_{K+Q}\right] \label{ll} \eeq where $K = (\omega,
\bk), Q = (\Omega,\bq)$, and, we remind the reader, $\int_K\equiv
T\sum_n\int d^dk/(2\pi)^d$. The ladder renormalizations within a
given particle-hole bubble can be absorbed into the
renormalization of one of Raman vertices: $\gamma^a (\bk) \sigma_3
\rightarrow \Gamma^a (\bk)$. The same also holds for the
renormalization of the density vertex: $\sigma_3 \rightarrow {\bar
\Gamma}^a(\bk)$. Replacing one of $\gamma^a (\bk)$ by
$\Gamma^a(\bk)$ in Eq. \ref{ll} and adding the series of RPA
renormalizations by $V_C$, as shown in Fig. \ref{fig:1}, we obtain
the full Raman susceptibility in the form
\begin{widetext}
\bea\label{eq:Ramannn2}
\chi_R (Q) &=&[-\pi_{RR}(Q)] + [-\pi_{RC}(Q)][-V_C(\bq)][-\pi_{CR}(Q)]+[-\pi_{RC}(Q)][-V_C(\bq)][-\pi_{CC}(Q)][-V_C(\bq)][-\pi_{CR}(Q)]+...\nonumber\\
&=&-\pi_{RR}(Q)-\pi_{RC}(Q) \frac{V_C(\bq)}{1-V_C(\bq)\pi_{CC}(Q)}\pi_{CR}(Q),\eea
\end{widetext}
where
\bea\label{eq:Raman2}
\pi_{RR}(Q)&=&\sum_{a}\int_K \text{Tr}\left[\gamma^a (\bk) \sigma_3 {G}^a_K {\Gamma}^a (\bk) {G}^a_{K+Q}\right],\\
\pi_{RC}(Q)&=&\sum_{a}\int_K \text{Tr}\left[\gamma^a (\bk) \sigma_3 {G}^a_K {\bar \Gamma}^a (\bk) {G}^a_{K+Q}\right], \label{eq:wtfm}\\
\pi_{CC}(Q)&=&\sum_{a}\int_K \text{Tr}\left[{\sigma_3} {G}^a_K
{\bar \Gamma}^a (\bk) {G}^a_{K+Q}\right]. \eea To evaluate
$\pi_{RR}(Q), \pi_{RC}(Q)$, and $\pi_{CC}(Q)$, we need the
expressions for the renormalized vertices $\Gamma^a (\bk)$ and
${\bar \Gamma}^a (\bk)$. The conventional way to obtain these
expressions  is to reduce the series of ladder diagrams for
$\Gamma^a$ and ${\bar \Gamma}_a$ to integral equations in
momentum, as schematically shown in Fig. \ref{fig:01}, and solve
these equations by expanding $\gamma^a (k)$, $\Gamma^a (k)$, and
${\bar \Gamma}_a (k)$ first in different irreducible
representations and then in eigenfunctions for  a given
irreducible representation. We shall refer to the various
components of this expansions as `partial components'. The partial
components from different irreducible representations decouple,
and the ones from the same irreducible representation form a set
of linear algebraic equations. One set  relates the prefactors for
partial components of  $\Gamma^a (k)$ to partial components of
$\gamma^a (k)$, the other relates the prefactors for partial
components of ${\bar \Gamma}_a (k)$ to the single non-zero partial
component of the bare density vertex, which does not depend on
$k$.  This procedure, however, can be implemented in Nambu
formalism  only if the interaction can be factorized as
$(\Psi^{\dag}_a(\bk) Q^{ab}\Psi_b(\bk+\bq)) \times
(\Psi^{\dag}_b(\bk'+\bq) Q^{ba} \Psi_a(\bk')$ with some matrices
$Q^{ab}$ and $Q^{ba}$.  The interactions $H_4$ and $H_1$ do have
these forms (with $Q^{ab} = Q^{ba} = \sigma_3$),  but $H_2$ and
$H_3$ do not, as evident from (\ref{eq:4_2}) and (\ref{eq:4_3}).
However, for the renormalization of the Raman bubble, we only need
parts of $H_2$ and $H_3$ which do have the required forms. To see
this we first observe that the Nambu matrix structure of the full
$\Gamma^a (\bk)$ and ${\bar \Gamma}_a (\bk)$ is \beq\label{ac_1}
\Gamma^a = \Gamma^a_3 \sigma_3 + \Gamma^a_2 \sigma_2, ~~{\bar
\Gamma}^a = {\bar \Gamma}^a_3 \sigma_3 + {\bar \Gamma}^a_2
\sigma_2. \eeq This structure can be verified by directly
evaluating the renormalized vertices in order-by-order
calculations. The $\sigma_3$ structure is present in the bare
vertices, and the renormalizations, which preserve it in the full
$\Gamma^a (\bk)$ and ${\bar \Gamma}_a (\bk)$, involve, in the
conventional Gorkov notation, the products of two normal and two
anomalous Greens functions in each cross-section.  We will be
calling these as renormalizations in the  ``particle-hole"
channel, because in the normal state they involve particle-hole
pairs of intermediate fermions. The $\sigma_2$ structure comes
from the processes which, in Gorkov notation, involve one normal
and one anomalous Green's function. Such processes transform a
particle-hole vertex into a particle-particle one. We will be
referring to these as renormalizations in the ``particle-particle"
channel. We next observe that the renormalizations in the
particle-hole channel involve interactions $V^{aa}_4$ and
$V^{ab}_2$ with the same spin projections for all four fermions,
while the ones in the particle-particle channel involve $V^{aa}_4$
and $V^{ab}_2$ with opposite spin projection for two pairs
fermions.  The corresponding  terms  in $H_2$ and $H_3$ are
\begin{widetext}
\bea
H_2 &\rightarrow& \sum_{a\neq b}\sum_{\bq} V^{ab}_2(q) \times \nonumber \\
&&\sum_{\bk,\bk'}\left[
\left(\Psi^{\dag}_a(\bk) \sigma_+ \Psi_b(\bk+\bq)\right) \left(\Psi^{\dag}_b(\bk'+\bq) \sigma_+\Psi_a(\bk')\right)\right] +
\left(\Psi^{\dag}_a(\bk) \sigma_- \Psi_b(\bk +\bq)\right) \left(\Psi^{\dag}_b(\bk'+\bq) \sigma_-\Psi_a(\bk')\right)
\label{eq:4_21} \\
H_3 &\rightarrow& \sum_{a\neq b}\sum_{\bq} V^{ab}_3(q) \times \nonumber \\
&& \sum_{\bk,\bk'} \left[\left(\Psi^{\dag}_a(\bk) \sigma_+
\Psi_b(\bk+\bq)\right) \left(\Psi^{\dag}_b(\bk'+\bq)
\sigma_-\Psi_a(\bk')\right) + \left(\Psi^{\dag}_a(\bk)
\sigma_-\Psi_b(\bk + \bq)\right) \left(\Psi^{\dag}_b(\bk'+\bq)
\sigma_+\Psi_a(\bk')\right)\right] \label{eq:4_31} \eea Both of
these terms have $(\Psi^{\dag}_a(\bk) Q^{ab}\Psi_b(\bk+\bq))
\times (\Psi^{\dag}_b(\bk'+\bq) Q^{ba} \Psi_b(\bk')$  Nambu matrix
structure. Using these forms, and the one for the  $V^{aa}_4$
interaction term, we obtain, after a simple algebra, the closed
set of equations \bea
{\Gamma}^a_3 (\bk) &=&\gamma^{a} (\bk) - \frac{1}{2} \int_K V^{aa}_4 (\bk-\bk') \text{Tr} \left[\sigma_3 \left(\sigma_3 {G}^a_{K'}\right)  \sigma_3  \left({G}^a_{K'+Q} \sigma_3\right)\right] {\Gamma}^a_3 (\bk ') \nonumber \\
&& -\frac{1}{2}\sum_{b \neq a}  \int_K  V^{ab}_2 (\bk-\bk') \text{Tr} \left[\sigma_3 \left(\sigma_+ {G}^b_{K'}\right) \sigma_3  \left({G}^b_{K'+Q} \sigma_+\right) +
\sigma_3 \left(\sigma_- {G}^b_{K'}\right) \sigma_3  \left({G}^b_{K'+Q} \sigma_-\right)\right]  {\Gamma}^b_3 (\bk ')  \nonumber \\
&&  - \frac{1}{2} \int_K  V^{aa}_4 (\bk-\bk') \text{Tr} \left[\sigma_3 \left(\sigma_3 {G}^a_{K'}\right)  \sigma_2  \left({G}^a_{K'+Q} \sigma_3\right)\right]  {\Gamma}^a_2 (\bk ')  \nonumber \\
&&-\frac{1}{2} \sum_{b \neq a}  \int_K  V^{ab}_2 (\bk-\bk') \text{Tr} \left[\sigma_3 \left(\sigma_+ {G}^b_{K'}\right) \sigma_2  \left({G}^b_{K'+Q} \sigma_+\right) +
\sigma_3 \left(\sigma_- {G}^b_{K'}\right) \sigma_2  \left({G}^b_{K'+Q} \sigma_-\right)\right]  {\Gamma}^b_2 (\bk ') \nonumber \\
{\Gamma}^a_2 (\bk) &=& - \frac{1}{2} \int_K  V^{aa}_4 (\bk-\bk') \text{Tr} \left[\sigma_2 \left(\sigma_3 {G}^a_{K'}\right)  \sigma_2  \left({G}^a_{K'+Q} \sigma_3\right)\right] {\Gamma}^a_2 (\bk ') \nonumber \\
&& -\frac{1}{2} \sum_{b \neq a}  \int_K  V^{ab}_3 (\bk-\bk') \text{Tr} \left[\sigma_2 \left(\sigma_+ {G}^b_{K'}\right) \sigma_2  \left({G}^b_{K'+Q} \sigma_-\right) +
\sigma_2 \left(\sigma_- {G}^b_{K'}\right) \sigma_2  \left({G}^b_{K'+Q} \sigma_+\right)\right]  {\Gamma}^b_2 (\bk ') \nonumber \\
 && - \frac{1}{2} \int_K  V^{aa}_4 (\bk-\bk') \text{Tr} \left[\sigma_2 \left(\sigma_3 {G}^a_{K'}\right)  \sigma_3  \left({G}^a_{K'+Q} \sigma_3\right)\right] {\Gamma}^a_3 (\bk ') \nonumber \\
  &&-\frac{1}{2} \sum_{b \neq a}  \int_K  V^{ab}_3 (\bk-\bk') \text{Tr} \left[\sigma_2 \left(\sigma_+ {G}^b_{K'}\right) \sigma_3  \left({G}^b_{K'+Q} \sigma_-\right) +
\sigma_2 \left(\sigma_- {G}^b_{K'}\right) \sigma_3  \left({G}^b_{K'+Q} \sigma_+\right)\right]  {\Gamma}^b_3 (\bk '),
 \label{vertex}
  \eea
and
\bea
{\bar \Gamma}^a_3 (\bk) &=&1 - \frac{1}{2}\int_K V^{aa}_4 (\bk-\bk') \text{Tr} \left[\sigma_3 \left(\sigma_3 {G}^a_{K'}\right)  \sigma_3  \left({G}^a_{K'+Q} \sigma_3\right)\right] {\bar\Gamma}^a_3 (\bk ')  \nonumber \\
&& -\frac{1}{2}\sum_{b \neq a}  \int_K  V^{ab}_2 (\bk-\bk') \text{Tr} \left[\sigma_3 \left(\sigma_+ {G}^b_{K'}\right) \sigma_3  \left({G}^b_{K'+Q} \sigma_+\right) +
\sigma_3 \left(\sigma_- {G}^b_{K'}\right) \sigma_3  \left({G}^b_{K'+Q} \sigma_-\right)\right]  {\bar \Gamma}^b_3 (\bk ')  \nonumber \\
&&  - \frac{1}{2} \int_K  V^{aa}_4 (\bk-\bk') \text{Tr} \left[\sigma_3 \left(\sigma_3 {G}^a_{K'}\right)  \sigma_2  \left({G}^a_{K'+Q} \sigma_3\right)\right]  {\bar \Gamma}^a_2 (\bk ')   \nonumber \\
&&-\frac{1}{2} \sum_{b \neq a}  \int_K  V^{ab}_2 (\bk-\bk') \text{Tr} \left[\sigma_3 \left(\sigma_+ {G}^b_{K'}\right) \sigma_2  \left({G}^b_{K'+Q} \sigma_+\right) +
\sigma_3 \left(\sigma_- {G}^b_{K'}\right) \sigma_2  \left({G}^b_{K'+Q} \sigma_-\right)\right]  {\bar \Gamma}^b_2 (\bk ') \nonumber \\
{\bar \Gamma}^a_2 (\bk) &=& - \frac{1}{2} \int_K  V^{aa}_4 (\bk-\bk') \text{Tr} \left[\sigma_2 \left(\sigma_3 {G}^a_{K'}\right)  \sigma_2  \left({G}^a_{K'+Q} \sigma_3\right)\right] {\bar \Gamma}^a_2 (\bk ')  \nonumber \\
&& -\frac{1}{2} \sum_{b \neq a}  \int_K  V^{ab}_3 (\bk-\bk') \text{Tr} \left[\sigma_2 \left(\sigma_+ {G}^b_{K'}\right) \sigma_2  \left({G}^b_{K'+Q} \sigma_-\right) +
\sigma_2 \left(\sigma_- {G}^b_{K'}\right) \sigma_2  \left({G}^b_{K'+Q} \sigma_+\right)\right]  {\bar \Gamma}^b_2 (\bk ') \nonumber \\
 && - \frac{1}{2} \int_K  V^{aa}_4 (\bk-\bk') \text{Tr} \left[\sigma_2 \left(\sigma_3 {G}^a_{K'}\right)  \sigma_3  \left({G}^a_{K'+Q} \sigma_3\right)\right] {\bar \Gamma}^a_3 (\bk ')  \nonumber \\
  &&-\frac{1}{2} \sum_{b \neq a}  \int_K  V^{ab}_3 (\bk-\bk') \text{Tr} \left[\sigma_2 \left(\sigma_+ {G}^b_{K'}\right) \sigma_3  \left({G}^b_{K'+Q} \sigma_-\right) +
\sigma_2 \left(\sigma_- {G}^b_{K'}\right) \sigma_3  \left({G}^b_{K'+Q} \sigma_+\right)\right]  {\bar \Gamma}^b_3 (\bk ').
 \label{vertex_c}
  \eea
\end{widetext}
Evaluating the traces over Pauli matrices,  we then obtain the set
of coupled integral equations in momentum for $\Gamma^a_{2,3}
(\bk)$ and ${\bar \Gamma}^a_{2,3} (\bk)$.  Each $\Gamma^a_{2,3}
(\bk)$ is expressed via $\int d^d k'$ over $\Gamma^b_{2,3} (\bk')$
with a kernel, that depends on $\bk'$ and $\bk-\bk'$. The same
holds for ${\bar \Gamma}^a_{2,3}$. We then separate different
irreducible lattice  representations, (e.g., one-dimensional
representations $A_{1g}, B_{1g}, B_{2g}, A_{2g}$ for the 2D square
lattice), and expand momentum-dependent interactions $V^{aa}_4
(\bk-\bk'), V_{2}^{ab} (\bk-\bk')$, $V_{3}^{ab} (\bk-\bk')$,  and
the vertices $\Gamma^a_{2} (\bk)$, $\Gamma^a_{3} (\bk)$, ${\bar
\Gamma}^a_{2} (\bk)$, ${\bar \Gamma}^a_{3} (\bk)$, into the set of
orthogonal eigenfunctions within a given representation as
\bea\label{eq:qwe}
&&V_4^{aa}(\bk-\bk')=\sum_{..} \sum_{tp} f^t_{\bk}V_4^{aa,tp} f^p_{\bk'},\nonumber\\
&&V_{2,3}^{ab} (\bk-\bk')= \sum_{..}\sum_{tp} f^t_{\bk}V_{2,3}^{ab,tp} f^p_{\bk'}, \nonumber \\
&&\Gamma^a_{2} (\bk)  \sum_{..} \sum_t \Gamma^{a,t}_{2}  f^t_{\bk},
~\Gamma^a_{3} (\bk) = \sum_{..} \sum_t \Gamma^{a,t}_{3}  f^t_{\bk},  \nonumber \\
&&{\bar \Gamma}^a_{2} (\bk) = \sum_{..} \sum_t {\bar \Gamma}^{a,t}_{2}  f^t_{\bk},
~{\bar \Gamma}^a_{3} (\bk) = \sum_{..} \sum_t {\bar
\Gamma}^{a,t}_{3}  f^t_{\bk}. \eea where $\sum_{..}$ stands for
the sum over representations. We also expand the Raman vertex
\beq\label{eq:eee} \gamma^a(\bk)=\sum_{..} \sum_t c^{a}_t
f^t_{\bk}. \eeq Substituting this into Eqs. (\ref{vertex}) and
(\ref{vertex_c}), separating the components, and using the fact
that eigenfunctions from different representations are orthogonal,
we obtain the set of equations for the partial components within a
given representation
\begin{widetext}
\bea
{\Gamma}^{a,t}_3 &=&c^{a}_t - \frac{1}{2} \int_K  \left[\sum_{p,m} \left(V^{aa,tp}_4 \left(\Pi^{aa,pm}_{32} \Gamma_2^{a,m} + \Pi^{aa,pm}_{33} \Gamma_3^{a,m}\right) +
\sum_b V^{ab,tp}_2 \left(\Pi^{bb,pm}_{32} \Gamma_2^{b,m} + \Pi^{bb,pm}_{33} \Gamma_3^{b,m}\right)\right)\right] \nonumber \\
{\Gamma}^{a,t}_2  &=& \frac{1}{2} \int_K  \left[\sum_{p,m} \left(V^{aa,tp}_4 \left(\Pi^{aa,pm}_{22} \Gamma_2^{a,m} + \Pi^{aa,pm}_{23} \Gamma_3^{a,m}\right) +
\sum_b V^{ab,tp}_3 \left(\Pi^{bb,pm}_{22} \Gamma_2^{b,m} + \Pi^{bb,pm}_{23} \Gamma_3^{b,m}\right)\right)\right] \nonumber \\
{\bar \Gamma}^{a,t}_3 &=&\delta_{t,1} - \frac{1}{2} \int_K  \left[\sum_{p,m} \left(V^{aa,tp}_4 \left(\Pi^{aa,pm}_{32} {\bar \Gamma}_2^{a,m} + \Pi^{aa,pm}_{33} {\bar \Gamma}_3^{a,m}\right) +
\sum_b V^{ab,tp}_2 \left(\Pi^{bb,pm}_{32} {\bar \Gamma}_2^{b,m} + \Pi^{bb,pm}_{33} {\bar \Gamma}_3^{b,m}\right)\right)\right] \nonumber \\
{\bar \Gamma}^{a,t}_2  &=& \frac{1}{2} \int_K  \left[\sum_{p,m} \left(V^{aa,tp}_4 \left(\Pi^{aa,pm}_{22} {\bar \Gamma}_2^{a,m} + \Pi^{aa,pm}_{23} {\bar \Gamma}_3^{a,m}\right) +
\sum_b V^{ab,tp}_3 \left(\Pi^{bb,pm}_{22} {\bar \Gamma}_2^{b,m} + \Pi^{bb,pm}_{23} {\bar \Gamma}_3^{b,m}\right)\right)\right]
\label{ac_2}
\eea
\end{widetext}
where \beq \Pi^{ab,pm}_{ij} =
\int_{K'}f^{p}_{\bk'}f^m_{\bk'}\text{Tr}\left[\sigma_i G^a_{K'}
\sigma_j G^b_{K'+Q}\right]. \eeq We remind the reader that $a,b$
label bands, $p,m$ label partial components, and $i,j=2,3$ label
the two sigma-matrices $\sigma_2$ and $\sigma_3$. Note that
$\Pi^{ab,pm}_{ij}$ with different $p$ and $m$ is generally
non-zero, because the corresponding eigenfunctions belong to the
same irreducible representation (e.g., $f^1_{\bk}=1, f^2_{\bk} =
\sqrt{2} \cos{4 \theta}$, etc for $A_{1g}$ representation in 2D,
at $|\bk| = k_F$). Eqs. (\ref{ac_2}) can be cast into the matrix
forms \bea\label{eq:eq}
\sum_{b,j,p}R^{ab,tp}_{ij}\Gamma_j^{b,p}=c^a_t\delta_{i,3}&\rightarrow&[\mathcal{R}] [\Gamma] = [c]\nonumber \\
\sum_{b,j,p}R^{ab,tp}_{ij}\bar\Gamma_j^{b,p}=\delta_{t,0}\delta_{i,3}&\rightarrow&[\mathcal{R}]
[\bar\Gamma] = [\bar c]. \eea Here $[\mathcal{R}]$ is the square
matrix with dimensions $2N_b n \times 2 N_b n$ where $n$ is the
number of components that we keep in Eq. (\ref{eq:qwe}). The
matrices $[\Gamma], [\bar\Gamma], [c]$, and $[\bar c]$ are vectors
with the dimension $2N_bn$.

The matrix $[\mathcal{R}]$ is determined by Eq. (\ref{ac_2}).  It
can be cast into the form: \beq
\mathcal{R}_{2N_Bn\times2N_Bn}=\left[1+ \frac12[V_{pp}][\Pi_{pp}]+
\frac12[V_{ph}][\Pi_{ph}]\right], \label{eq:Ram4ab} \eeq where
\begin{widetext}
\bea\label{ac_3}
[V_{pp}] = \left(\begin{array}{ccccccc}
V^{11,11}_4&V^{12,11}_{3}...&V^{1N_b,11}_{3}&V^{11,12}_4&V^{12,12}_{3}...&V^{1N_b,12}_{3}&...\\
V^{21,11}_3&V^{22,11}_{4}...&V^{2N_b,11}_{3}&V^{21,12}_3&V^{22,12}_{4}...&V^{2N_b,12}_{3}&...\\
&&&.&&&\\
&&&.&&&\\
&&&.&&&\\
V^{Nb1,n1}_3&V^{N_b2,n1}_{3}...&V^{N_bN_b,n1}_{4}&V^{N_b1,n2}_3&V^{22,n2}_{3}...&V^{N_bN_b,n2}_{4}&...
\end{array}\right)_{N_bn\times N_bn}\otimes\mathbb{1}_{2\times2}
\eea and  $[V_{ph}]$ has the same form as $[V_{pp}]$ with
$V_3\rightarrow V_2$. The matrix $[\Pi_{pp}]$ is  given by
\beq\label{abc_3} [\Pi_{pp}] = \left(\begin{array}{ccccccc}
P^{11,11}&0_{2\times2}...&0_{2\times2}&P^{11,12}&0_{2\times2}...&0_{2\times2}&...\\
0_{2\times2}&P^{22,11}...&0_{2\times2}&0_{2\times2}&P^{22,12}...&0_{2\times2}&...\\
&&&.&&&\\
&&&.&&&\\
&&&.&&&\\
0_{2\times2}&0_{2\times2}...&0_{2\times2}&P^{N_bN_b,n1}&0_{2\times2}...&P^{N_bN_b,n2}&...
\end{array}\right)_{N_bn\times N_bn},~~~~~~~~~~~~~~~~~~~~~~\eeq
\end{widetext}
where
\beq
P^{aa,mt}=\left(\begin{array}{cc}
-\Pi^{aa,mt}_{22}&-\Pi^{aa,mt}_{23}\\
0&0
\end{array}
\right),~
0_{2\times2} =\left(\begin{array}{cc}
0&0\\
0&0
\end{array}
\right). \eeq The matrix $[\Pi_{ph}]$ is the same as $[\Pi_{pp}]$,
with $P\rightarrow \tilde P$, where \beq\label{eq:abcd} \tilde
P^{aa,mt}=\left(\begin{array}{cc}
0&0\\
\Pi^{aa,mt}_{32}&\Pi^{aa,mt}_{33}
\end{array}
\right).
\eeq
Finally,
\bea\label{eq:c}
[c]^T&=&(C^1_1,C^2_1,..C^{N_b}_1,C^{1}_2,...C^{N_b}_2,...);\nonumber\\
C^a_t&=&(0,c^a_t),
\eea
and
\bea\label{eq:bc}
[\bar c]^T&=&(\bar C^1_1,\bar C^2_1,..\bar C^{N_b}_1,0,...0,...);\nonumber\\
\bar C^a_1&=&(0,1).
\eea
Using the above expressions we obtain
\bea
\pi_{RR}&=&\sum_{a}\sum_{t_1t_2}\sum_{j=2,3}
c_{t_1}^a\Pi^{aa,t_1t_2}_{3j}{\Gamma}^{a,t_2}_{j},\label{eq:fan55}\\
\pi_{RC}&=&\sum_{a}\sum_{t_1t_2}\sum_{j=2,3}
c_{t_1}^a\Pi^{aa,t_1t_2}_{3j}{\bar\Gamma}^{a,t_2}_{j},\label{eq:fan56}\\
\pi_{CR}&=&\sum_{a}\sum_{t_1t_2}\sum_{j=2,3}
\delta_{t_1,1}\Pi^{aa,t_1t_2}_{3j}{\Gamma}^{a,t_2}_{j},\label{eq:fan57}\\
\pi_{CC}&=&\sum_{a}\sum_{t_1t_2}\sum_{j=2,3}
\delta_{t_1,1}
\Pi^{aa,t_1t_2}_{3j}{\bar\Gamma}^{a,t_2}_{j}.\label{eq:fan58} \eea
All these quantities depend on $Q = (\Omega, \bq)$ via the
polarization operators. The quantities $\pi_{CR}$ and $\pi_{RC}$
are indeed equal. Eqs. (\ref{eq:fan55}) - (\ref{eq:fan58})
together with the Eq. (\ref{eq:Ramannn2}) relating $\pi_{RR},
\pi_{RC} = \pi_{CR}$ and $\pi_{CC}$ to  Raman susceptibility
$\chi_{R}(\Omega, \bq)$ comprise the general formula for the Raman
intensity $R(\Omega) \propto \text{Im} \chi_{R} (\Omega, q\to 0)$.
These relations are valid for any number of bands, any pairing
symmetry, and any Raman scattering geometry. Raman-active
collective modes show up as poles in $\chi_{R} (\Omega, \bq)$ and
spikes in $R(\Omega)$.

In the next two sections (Sec. \ref{sec:ex} and Sec.
\ref{sec:screening}), we  present a case-by-case analysis of the
$A_{\rm 1g}$ Raman intensity at $T=0$ in one-band and two-band 2D
$s-$wave SCs on a square lattice. In Sec. \ref{sec:ex}, we
investigate the response  without Coulomb interaction, i.e.
approximate $\chi_R (\Omega)$ by $-\pi_{RR} (\Omega)$.  We
consider separately the effects due to renormalizations of the
Raman bubble in the particle-particle and particle-hole channels.
In Sec. \ref{sec:screening} we discuss the contribution to Raman
response from Coulomb interaction.

\section{Application to A$_{\rm 1g}$ channel}\label{sec:ex}
It is clear that from Eqs. \ref{eq:fan55}-\ref{eq:fan58} that
essential quantities needed for the Raman response Fare the
polarization bubbles $\Pi_{ij}^{aa,mt}$ for various harmonics
belonging to the A$_{\rm 1g}$ representation: $\{1, \cos k_x+\cos
k_y,...\}$ over the BZ, or $f_{\bk}
=\{1,\cos4\theta,\cos8\theta,...\}$ over the Fermi-surface, where
$\theta$ is the angle which $\bk$ makes with the $k_x$ axis in the
BZ. In general, the pairing gap $\Delta$ and the density of states
on the Fermi surface $\nu_F$ also contain infinite number of
$A_{1g}$ components. For the sake of transparency, we assume that
$\Delta$ and the density of states $\nu_F$ are isotropic. These
assumptions are made only to simplify the presentation and be able
to compute $\chi_{R} (\Omega)$ analytically.

\subsection{One-band $s$-wave SC}\label{sec:i1bandex}
We start with the one-band case -- one FS, centered at the
$\Gamma$-point. This case has been analyzed diagrammatically by
Klein and Dierker,\cite{Klein} and we indeed reproduce their
results.  In contrast to Ref. \onlinecite{Klein}, however, we
analyze the effects of short-range and Coulomb interactions
separately. We  show that there is a strong reduction [and full
cancellation for $\gamma (\bk) =$const] of the $A_{1g}$ Raman
response already due to vertex corrections in the
particle-particle channel. This was not emphasized in Ref.
[\onlinecite{Klein}], although it follows from the formulas
presented in that work.  The reduction/cancellation of $A_{\rm
1g}$ response due to vertex corrections has been demonstrated in
Ref. [~\onlinecite{Lara}], where the authors used effective action
approach rather than direct diagrammatics.

\subsubsection{Isotropic case}
In an isotropic case, there is only one component of  $f_{\bk}$
and $\gamma_{\bk}$ in A$_{\rm 1g}$ geometry: $f_{\bk} =1,
\gamma_{\bk}  = c_1$. For the one-band model, the only interaction
is $V^{11,11}_4 = V_4$, and we take it to be attractive, i.e.,
$V_4 <0$.  To compute  the Raman response, we need the expressions
for four $2 \times 2$ matrices  $\Pi_{ij}^{11,11} = \Pi_{ij}$ with
$i,j =2,3$. Evaluating $\Pi_{ij} = \int_{K'} Tr\left[\sigma_i
G^1_{K'} \sigma_j G^1_{K' + Q}\right]$, we obtain \beq
[V_{pp}]=V_4\sigma_0;~[\Pi_{pp}]=\left(\begin{array}{cc}
-\Pi_{22}&-\Pi_{23}\\
0&0
\end{array}
\right)\nonumber
\eeq
\beq\label{eq:ddd}
[V_{ph}]=V_4\sigma_0;~[\Pi_{ph}]=\left(\begin{array}{cc}
0&0\\
\Pi_{32}&\Pi_{33}\\
\end{array}
\right),
\eeq
where, as before, $\sigma_0$ is the $2\times 2$ unity matrix, and
\bea\label{eq:123}
\Pi_{22}(\Omega)&=& \frac{2}{V_4}-\left(\frac{\Omega}{2\Delta}\right)^2F(\Omega),\nonumber\\
\Pi_{23}(\Omega)&=&\frac{i\Omega}{2\Delta}F(\Omega),\nonumber\\
\Pi_{32}(\Omega) &=&  - \Pi_{23}(\Omega),\nonumber\\
\Pi_{33}(\Omega)&=&-F(\Omega),
\eea
where
\bea
F(\Omega)&=&\int_{\bk} \frac{\Delta^2}{E_{\bk}\left[E^2_{\bk}-(\Omega/2)^2\right]} \nonumber \\
&=&  2\nu_F\frac{\sin^{-1}\left(\Omega/2\Delta\right)}
{(\Omega/2\Delta)\sqrt{1-\left(\Omega/2\Delta\right)^2}} \nonumber \\
E_{\bk} &=& \sqrt{[\epsilon (\bk)]^2 + \Delta^2}, \eea In Eq.
(\ref{eq:123}) we used self-consistency condition on $\Delta$, Eq.
(\ref{eq:self}) at $T=0$ which yields \beq
\int_{\bk}\frac{1}{2E_{\bk}}=-\frac{1}{V_4}. \eeq The matrix
$[c]^T$ is $(0,c_1)$, and \beq
\mathcal{R}\equiv\left(\begin{array}{cc}
\mathcal{R}_{22}&\mathcal{R}_{23}\\
\mathcal{R}_{32}&\mathcal{R}_{33}
\end{array}
\right)=\sigma_0+\frac12[V_{pp}][\Pi_{pp}]+\frac12[V_{ph}][\Pi_{ph}] \label{eq:xxx}
\eeq
\beq \label{eq:a21}
~~~~~=\left(\begin{array}{cc}
1-\frac{V_4}{2}\Pi_{22}&-\frac{V_4}{2}\Pi_{23}\\
\frac{V_4}{2}\Pi_{32}&1+\frac{V_4}{2}\Pi_{33}
\end{array}\right).
\eeq
Calculating $[\Gamma]$ as $[\mathcal{R}]^{-1}[c]$ from Eq. (\ref{eq:eq}) and using Eq. (\ref{eq:fan55}), we see that the Raman response $\chi_R(\Omega)$ is given by
\bea \label{aa}
\chi_R(\Omega)&=&(c_1)^2
\frac{\Pi_{32}\mathcal{R}_{23}-\Pi_{33}\mathcal{R}_{22}}{\text{Det}[\mathcal{R}]} \nonumber \\
&=& - (c_1)^2\left[\frac{\Pi_{33}-\frac{(\Pi_{23})^2}{2/V_4-\Pi_{22}}}
{1+\frac{V_4}{2}\left(\Pi_{33}-\frac{(\Pi_{23})^2}{2/V_4-\Pi_{22}}\right)}\right]
\eea

\subsubsection{Role of $V_{pp}$}
Let us first analyze only vertex corrections in the particle-particle channel. To do this, we momentarily set
$[V_{ph}]$ term in Eq. \ref{eq:Ram4ab} to zero. We denote the corresponding Raman response as  $\chi_R^{pp}(\Omega)$. We obtain
\beq\label{eq:ee}
\chi_R^{pp}(\Omega)= - (c_1)^2\left\{\Pi_{33}-\frac{(\Pi_{23})^2}{\frac{2}{V_4}-\Pi_{22}}\right\}
\eeq
Substituting the expressions for $\Pi_{ij}$ from Eq. (\ref{eq:123}), we find that $\chi_R^{pp}(\Omega)$ vanishes:
\bea\label{eq:324}
\chi_R^{pp} (\Omega)&=&(c_1)^2\left\{F(\Omega) - \frac{(\Omega/2)^2F^2(\Omega)}{\frac{2}{V_4}- \frac{2}{V_4}+(\Omega/2)^2F(\Omega)}\right\}\nonumber\\
&=&0.
\eea
The first $F(\Omega)$  term in Eq. (\ref{eq:324}) is a free-fermion particle-hole polarization bubble. Taken alone, this term would give rise to $2\Delta$ singularity in the Raman response and to non-zero Raman intensity $\propto  \text{Im}[\chi^{pp}_R (\Omega)]$  at $\Omega > 2\Delta$.  The second term, that cancels $F(\Omega)$, is the contribution from vertex corrections in the particle-particle channel. The cancellation of the two $\frac{2}{V_4}$ in the denominator of Eq. (\ref{eq:324}) is guaranteed by the U(1)-gauge invariance: the Raman susceptibility  must contain the pole corresponding to the BAG mode, and the vanishing of the denominator in Eq. (\ref{eq:324}) at $\Omega =0$ ensures that this mode is massless. The vanishing of $\chi^{pp}_R (\Omega)$ at all frequencies is the consequence of the conservation of the number of fermions (or, equivalently, of the total charge). Indeed, for $\gamma_{\bk} = c_1$, the Raman vertex becomes identical to the density vertex, and particle conservation imposes that the density-density bubble must vanish at  $\bq\rightarrow0$ and any finite $\Omega$. We emphasize that this vanishing holds independent of whether we include long-range Coulomb interaction.

Despite the vanishing of $\chi^{pp}({\Omega})$  is expected  in an
isotropic case on general grounds, it has not been discussed in
Raman literature until recently.\cite{Lara} Several
authors\cite{dev_2,Deveaux,Klein,PeterBoyd,Girsh} presented
results for A$_{\rm 1g}$ Raman intensity that vanishes due to
screening by long-range Coulomb interaction if $\gamma(\bk)$ is
treated as a constant. Our analysis (and the one in Ref.
\onlinecite{Lara}) shows that  the $A_{1g}$ Raman intensity in an
$s-$wave SC vanishes in the isotropic case already before one
includes long-range Coulomb interaction. The physics argument is
that the original, normal state Hamiltonian with four-fermion
interaction $V_{pp}$  conserves the number of particles, hence
once all effects due to $V_{pp}$ are included (i.e., the
contributions from the superconducting condensate {\it and} from
renormalizations in the particle-particle channel within the
particle-hole bubble), the fully dressed density-density bubble
should obey the same conservations laws as in the normal state,
i.e., it should vanish at $q=0$ and finite $\Omega$.
\begin{figure}[htp]
$\begin{array}{c}
\includegraphics[width=0.9\columnwidth]{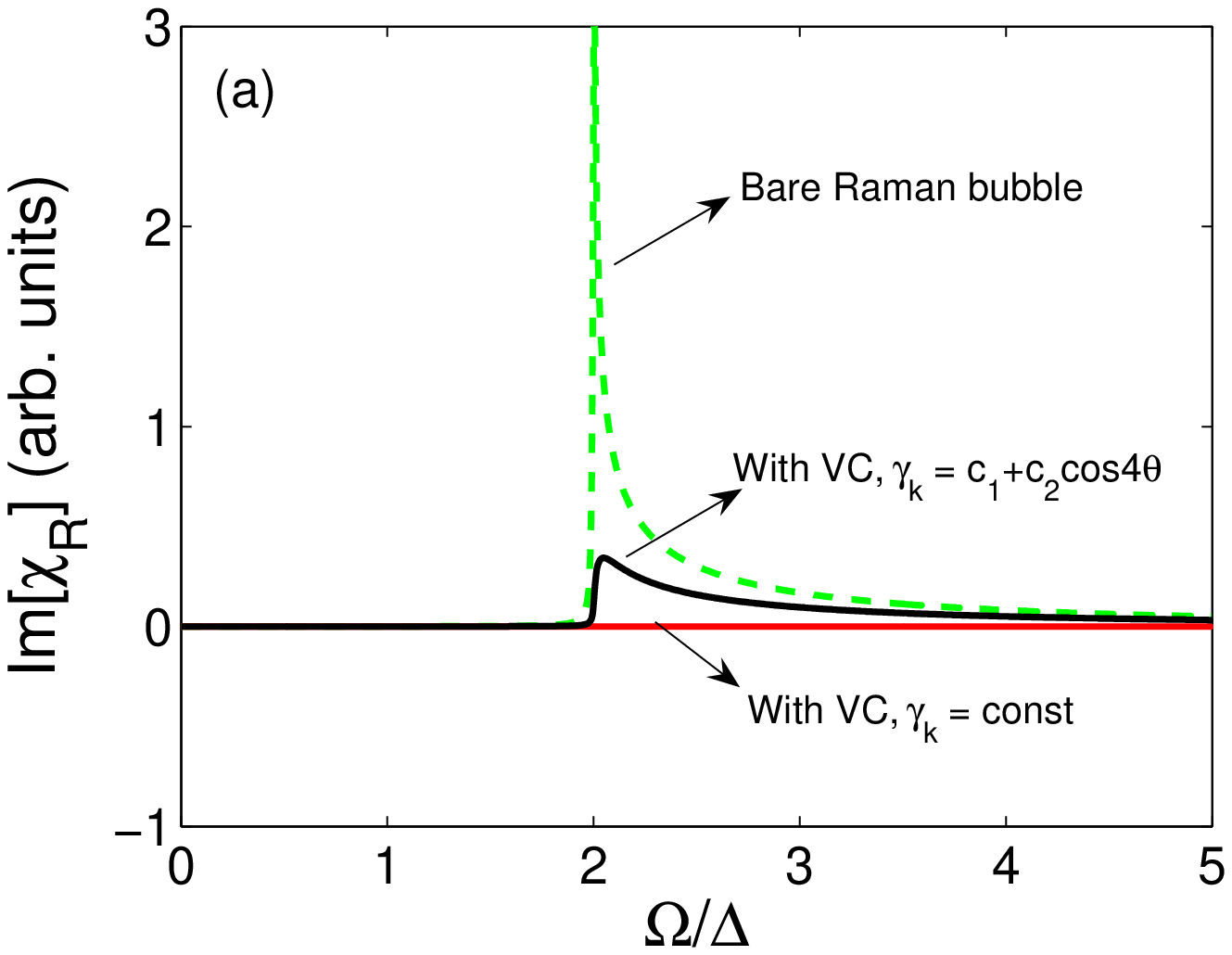}\\
\includegraphics[width=0.9\columnwidth]{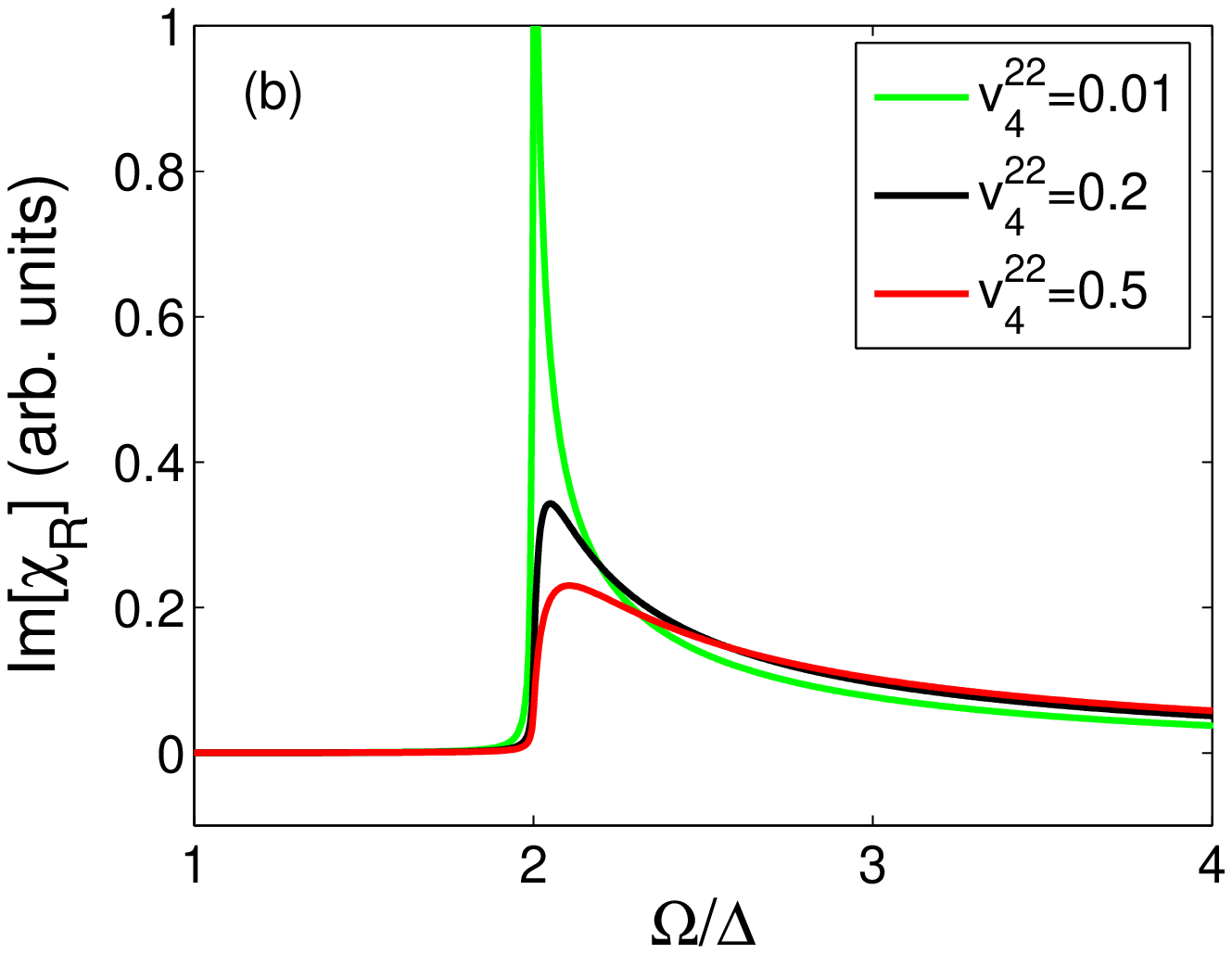}\\
\end{array}$
\caption{
\label{fig:11} Color online: (a) The Raman response, for a 1-band $s-$wave SC in the A$_{1g}$ channel, where it is approximated by the bare bubble contribution (dashed green line), the vertex corrected (VC) contribution with a constant $\gamma_{\bk}$ (red line) and VC contribution with a momentum dependent $\gamma_{\bk}$ (black line). (b) The removal of the $2\Delta$ edge singularity by vertex corrections from the subleading channel interaction $v^{22}_{4}\equiv \nu_FV_4^{22}$. The peak shifts to larger $\Omega$ as the interaction strength in increased. We have used $V^{11}_4=-0.3/\nu_F$, $c_1=0.3,~c_2=0.2$. A fermion lifetime of $0.005\Delta$ was added to obtain the broadening.}
\end{figure}

\subsubsection{Role of $V_{ph}$}
We now show that $A_{1g}$  Raman intensity in the isotropic case still vanishes, even if we include the renormalizations in the particle-hole channel. Indeed, comparing Eqs.
 (\ref{aa}) and (\ref{eq:ee}) we immediately find that
\bea\label{eq:22_1}
\chi_R(\Omega)&=&  \frac{\chi_R^{pp}(\Omega)}
{1- \frac{V_4}{2c^2_1} \chi_R^{pp} (\Omega)}
\eea
Because $\chi_R^{pp}(\Omega) =0$, the full Raman response, with particle-particle {\it and} particle-hole vertex corrections, also vanishes.   This is indeed expected because $V_{ph}$ preserves the number of particles.

\subsection{Case of anisotropic Raman vertex for one-band SC}\label{sec:a1bandex}

We now consider the case when the Raman vertex  has two partial components (from the same irreducible representation):  $\gamma_{\bk} = c_1 f^1_{\bk} + c_2 f^2_{\bk}$. For definiteness we take
$f^1_{\bk} =1, f^2_{\bk} = \sqrt{2} \cos{4\theta}$.
We use  Eq. (\ref{eq:qwe}) to decompose $V_4 (\bq) = V_4 (\bk-\bk')$ into two harmonics
\bea
V_4 (\bk-\bk')&=&f^1_{\bk} V_{4}^{11}f^1_{\bk'} + f^2_{\bk} V_{4}^{22} f^2_{\bk'} +\nonumber\\
&&f^1_{\bk} V_{4}^{12} f^2_{\bk'} +  f^2_{\bk} V_{4}^{21} f^1_{\bk'}  \nonumber \\
&=&V_{4}^{11} + 2 V_{4}^{22} {\cos{4\theta_k}}  {\cos{4\theta_{k'}}} + \nonumber\\
&&\sqrt{2} V_4^{12} \left( {\cos{4\theta_k}} +  {\cos{4\theta_{k'}}}\right).
\eea
The off-diagonal terms can be eliminated by a rotation to new eigenfunctions,\cite{SMPJH} which are linear combinations of a constant and ${\cos{4\theta_k}}$.  We will not do this, but just set here $V_4^{12}= V_4^{21} =0$. We will discuss a more generic case in Sec. V, when we analyze the role of Coulomb interaction. As before, we assume that SC is induced by the interaction $V_4^{11}$, which we keep negative (attractive). The corresponding gap $\Delta$ is then isotropic. The interaction $V_4^{22}$ can be either repulsive or attractive. In the latter case, we assume that it is weaker than $V_4^{11}$.

The matrices $[V_{pp}]$ and $[V_{ph}]$ now become $4\times 4$ matrices ($N_b=1, n=2$, $2N_b n =4$).  We have (dropping the band indices $ab$, i.e., setting
$V^{ab, mt}_i = V^{mt}_i, \Pi_{ij}^{ab,mt} = \Pi_{ij}^{mt}$)
\beq\label{eq:ses}
[V_{pp}]=\left(\begin{array}{cc}
V_4^{11} \sigma_0&0\\
0& V_4^{22} \sigma_0
\end{array}
\right);~[V_{ph}]=\left(\begin{array}{cc}
V_4^{11}\sigma_0&0\\
0&V_4^{22}\sigma_0
\end{array}
\right);\nonumber
\eeq
The matrices $[\Pi_{pp,ph}]$ become
\beq\label{eq:ses2}
[\Pi_{pp}]=\left(\begin{array}{cc}
P^{11}&0\\
0&P^{22}
\end{array}
\right);~[\Pi_{ph}]=\left(\begin{array}{cc}
\tilde{{P}}^{11}&0\\
0&\tilde{{P}}^{22}
\end{array}
\right);\nonumber
\eeq
where
\beq
{P}^{mm} =\left(\begin{array}{cc}
-\Pi^{mm}_{22}&-\Pi^{mm}_{23}\\
0&0
\end{array}
\right),~~
\tilde{{P}}^{mm} = \left(\begin{array}{cc}
0&0\\
\Pi^{mm}_{32}&\Pi^{mm}_{33}\\
\end{array}
\right),
\label{bb_1}
\eeq
 $m\in\{1,2\}$, and
\beq\label{eq:ses3}
[c]^T=\left(0,c_1,0,c_2\right).
\eeq
Substituting this into Eq (\ref{eq:xxx}) we obtain
\begin{widetext}
\beq\label{eq:sees}
\mathcal{R}=\left(\begin{array}{cccc}
1-\frac{V_4^{11}}{2}\Pi^{11}_{22}&-\frac{V_4^{11}}{2}\Pi^{11}_{23}&0&0\\
\frac{V_4^{11}}{2}\Pi^{11}_{32}&1+\frac{V_4^{11}}{2}\Pi^{11}_{33}&0&0\\
0&0&1-\frac{V_4^{22}}{2}\Pi^{22}_{22}&-\frac{V_4^{22}}{2}\Pi^{22}_{23}\\
0&0&\frac{V_4^{22}}{2}\Pi^{22}_{32}&1+\frac{V_4^{22}}{2}\Pi^{22}_{33}
\end{array}
\right);\nonumber
\eeq
For angle-independent $\Delta$,  $\Pi^{11}_{ij}=\Pi^{22}_{ij} = \Pi_{ij}$. Inverting the matrix $\mathcal{R}$ and using Eq. (\ref{eq:fan55}) we get
\bea\label{eq:sudden}
\chi_R(\Omega)&=&-(c_1)^2\left[\frac{\Pi_{33}-\frac{(\Pi_{23})^2}{2/V_4^{11}-\Pi_{22}}}
{1+\frac{V_4^{11}}{2}\left(\Pi_{33}-\frac{(\Pi_{23})^2}{2/V_4^{11}-\Pi_{22}}\right)}\right] - (c_2)^2\left[\frac{\Pi_{33}-\frac{(\Pi_{23})^2}{2/V_4^{22}-\Pi_{22}}}
{1+\frac{V_4^{22}}{2}\left(\Pi_{33}-\frac{(\Pi_{23})^2}{2/V_4^{22}-\Pi_{22}}\right)}\right],
\eea
\end{widetext}
The term with the prefactor $(c_1)^2$ vanishes, as in the isotropic case, but  the term with  the prefactor $c_2^2$ remains finite when $V_4^{22} \neq V_4^{11}$. As a result, the Raman response $R(\Omega)\propto\text{Im}[\chi_R(\Omega)]$ is non-zero:
\bea\label{eq:eew_1}
R(\Omega)&\propto & (c_2)^2 \text{Im} \left[\frac{\Pi_{33}-\frac{(\Pi_{23})^2}{2/V_4^{22}-\Pi_{22}}}{1+\frac{V_4^{22}}{2}\left(\Pi_{33}-\frac{(\Pi_{23})^2}{2/V_4^{22}-\Pi_{22}}\right)}\right],\nonumber\\
\eea
Because $F(\Omega)$ is real at $\Omega <2\Delta$, $\Pi_{33}, (\Pi_{23})^2$, and $\Pi_{22}$ are also real. Then $R(\Omega<2\Delta) =0$, except for $\Omega$ at which the denominator in (\ref{eq:eew_1}) vanishes. At such frequencies $R(\Omega)$ has $\delta$-function peaks.  Whether such peaks exist depends on the sign and magnitude of $V_4^{22}$.  For attractive  $V_4^{22} <0$ a simple analysis shows that the denominator in (\ref{eq:eew_1}) does vanish at
\beq\label{eq:qq}
1 - \frac{V_4^{22}}{V_4^{11}} = \left(\frac{\Omega}{2\Delta}\right)^2 \frac{F(\Omega)}{\frac{2}{|V_4^{22}|} + F(\Omega)}
\eeq
We recall that we required $|V^{22}_4|/|V^{11}_{4}|<1$. Then the left hand side of Eq. (\ref{eq:qq}) is less than unity for an attractive $V_4^{22}$ (recall that $V_4^{11}<0$ for SC state to exist). Since $F(0)$ is a constant and $F(2\Delta-0)$ diverges, we  immediately see that the right hand side of Eq. (\ref{eq:qq}) ranges from 0 to 1,
i.e., this equation necessarily has a solution at some $\Omega < 2\Delta$. At such a frequency $\chi_R (\Omega)$ has a pole.  Because the pole emerges only for  $V_4^{22} <0$,  it is tempting to  associate it with the  BS-type mode (i.e., oscillations of the pairing order parameter in the secondary attractive channel). Note, however, that the true BS-type mode would be at a frequency where $2/V_4^{22} = \Pi_{22} (\Omega)$ (see Ref. \onlinecite{SMPJH}). [The `original' BS mode is in the $d$-wave channel. Here we refer to `BS-type modes', whose symmetry is associated with the same irreducible representation as the condensate].  In our case,  the position of the $\delta-$function peak in $R(\Omega)$ is shifted from this frequency  due to renormalizations in the particle-hole channel.  Note that for positive $V^{22}_4$, there is no BS-type mode, i.e., no solution of Eq. (\ref{eq:qq}) along real frequency axis.

We now analyze what happens near  $\Omega = 2\Delta$.  Here $F(\Omega)\approx i2\nu/\sqrt{2x}$, where $x = \Omega/(2\Delta) -1$.  Using Eq. (\ref{eq:123}) we then obtain
\beq\label{eq:wqe}
R(\Omega) \propto \frac{(2/V_4^{22}-2/V_4^{11})^2}{2\nu_F} \text{Re} \left[\sqrt{2x}\right].
\eeq
If we only kept $\Pi_{33}$ term in Eq. (\ref{eq:eew_1}), we would obtain $1/\sqrt{x}$ singularity at $x = 0+$. We see that vertex corrections force the response at $\Omega =2\Delta +0$, ($x=0+$) to be zero. This effect was first pointed out in Ref. \onlinecite{zawadowskii} where the authors
discussed the collective mode contribution to the Raman intensity in the B$_{\rm 1g}$ channel. As $\Omega$  increases above $2\Delta$,  $R(\Omega)$ increases.  At very large $\Omega$, $R(\Omega)$ tends to zero, as in this limit the system recovers normal state behavior, where $R(\Omega)$ vanishes within our approximation.  In between, $R(\Omega)$ passes through a maximum at some $\Omega > 2\Delta$.  The location of the maximum depends on the relative values  of $\nu_F |V_{4}^{22}|$ and $\nu_F |V_{4}^{11}|$. When both are small and not too close to each other, the deviations of the functional form of $R (\Omega)$ from that of $-\text{Im}[\Pi_{33} (\Omega)] = \text{Im}[F(\Omega)]$ become essential only near $\Omega =2\Delta$, when $\sqrt{x} \leq  (|V_{4}^{11}| \nu_F) *(|V_4^{22}|\nu_F)/(|V_{4}^{11}| \nu_F) -(|V_4^{22}|\nu_F)$. For larger $x$, $R(\Omega)$ has the same $1/\sqrt{x}$ behavior as $\text{Im}[F(\Omega)]$.

Figs.\ref{fig:11}(a) and (b) summarize our results for the one-band case. Figure (a) shows how the vertex corrections remove the $2\Delta$ edge singularity (and even force the response to be zero if $\gamma_{\bk}$ is constant). Figure (b) shows shifting of the ``$2\Delta$-peak" to higher $\Omega$ as a result of vertex corrections from the interaction in the subleading channel that doesn't contribute to the pairing.

\subsection{Case of isotropic two-band system}\label{sec:i2bandex}
We now extend the analysis  to a two-band SC. For definiteness we consider two pockets ($a$ and $b$) around the $\Gamma$-point. We start with the isotropic case, when $\gamma^a (\bk) = c^a_{1}$, $\gamma^b (\bk) = c^b_{1}$. Then we only have to include momentum-independent components of the interaction $V_i^{ab,11} \equiv V_i^{ab}$.   At the same time, for two bands we have to include three types of interactions, $V_i^{ab}$ with $i =2,3,4$ (see Sec.  \ref{sec:Raman}). To simplify the notations, we set
\beq
V^{aa}_4 = V_a, V^{bb}_4 = V_b, V_2^{ab} = V_2, V_3^{ab} = V_3
\eeq
The pairing gaps $\Delta_a$ and $\Delta_b$ are determined by the interplay between intra-band interactions $V_a$ and $V_b$ and the inter-band pair-hopping interaction $V_3$ (because we pair electrons in the same band, $V_2$ does not appear here):
\beq\label{eq:Ds2019}
\left(\begin{array}{c}
\Delta^a\\
\Delta^b
\end{array}
\right)=-
\left(\begin{array}{cc}
V_a&V_3\\
V_3&V_b
\end{array}
\right)\left(\begin{array}{c}
\Delta^a l^a\\
\Delta^b l^b
\end{array}
\right),
\eeq
where $l^a=\int_{\bk}1/2E^a_{\bk}$. We consider  the case when the pairing is due to intra-band attraction ($V_a$, $V_b <0$, $V_a V_b > V^2_3$),  and when it is due to inter-band interaction, $V_a V_b <V^2_3$. In the second case, one can easily obtain from Eq. (\ref{eq:Ds2019}) that $V_3\Delta^a\Delta^b$ is negative, i.e., $s-$wave SC is of $s^{++}$ type when $V_3 <0$ and of $s^{+-}$ type when $V_3 >0$. The interaction $V_2$ does not contribute to the pairing or to vertex renormalizations in the particle-particle channel, but it contributes to vertex renormalizations in the particle-hole channel.

\begin{figure}[htp]
\includegraphics[width=0.8\columnwidth]{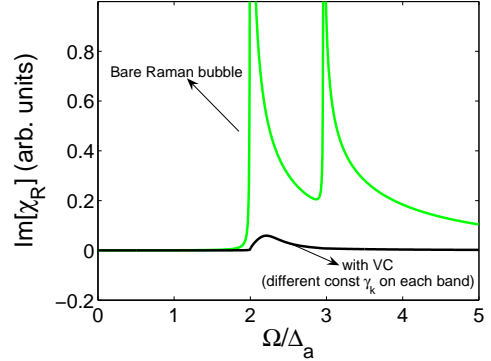}
\caption{
\label{fig:ex}  The Raman response of a 2-band $s$-wave SC. The light line is the result when one includes only the bare bubble, dark line is the full result, with vertex corrections(VC) included.
These corrections remove the edge singularity and move the peak to a frequency larger than $2\Delta_{\rm min}= \Delta_a$. The Raman vertex for each band is chosen to be a constant with $c^a_1=0.3$ and $c^b_1=0.2$. Here $V_a=-0.2/\nu_F,~V_b=-0.3/\nu_F,~V_3=0.1/\nu_F$, and $\nu_{F,a}=\nu_{F,b}=\nu_{F}$.}
\end{figure}

To calculate the Raman susceptibility $\chi_R (\Omega) = -\pi_{RR} (\Omega)$, we will need
\bea\label{eq:ds}
\Pi^{aa}_{23}&=&\frac{i\Omega}{2\Delta^a}F_a,\nonumber\\
\Pi^{aa}_{33}&=&-F_a,\nonumber\\
\text{where}~~F_a&=&\int_{\bk}\frac{(\Delta^a)^2}{E^a_{\bk}\left[(E^a_{\bk})^2-\left(\Omega/2\right)^2\right]} \nonumber \\
&& =  2\nu_{F,a} \frac{\sin^{-1}\left(\Omega/2\Delta^a\right)}
{(\Omega/2\Delta^a)\sqrt{1-\left(\Omega/2\Delta^a\right)^2}} \nonumber \\
\text{and}~~E^a_{\bk} &=& \sqrt{[\epsilon^a (\bk)]^2 + \Delta_a^2},
\eea
We also need
\beq
\Pi^{aa}_{22}=-2l^a-\left(\frac{\Omega}{2\Delta^a}\right)^2 F_a.
\eeq
These expressions are used to construct $[\Pi_{pp,ph}]$  defined in Eq. (\ref{abc_3}). In the isotropic two-band case $[c]^T=(0,c_1^a,0,c_1^b)$, and from Eq. (\ref{eq:Ram4ab})
\begin{widetext}
\bea\label{eq:21}
\mathcal{R}&=&
\mathbb{1}_4 +\frac12\left(\begin{array}{cccc}
V_a&0&V_3&0\\
0&V_a&0&V_3\\
V_3&0&V_b&0\\
0&V_3&0&V_b
\end{array}\right)\left(\begin{array}{cccc}
-\Pi^{aa}_{22}&-\Pi^{aa}_{23}&0&0\\
0&0&0&0\\
0&0&-\Pi^{bb}_{22}&-\Pi^{bb}_{23}\\
0&0&0&0
\end{array}\right)+\frac12\left(\begin{array}{cccc}
V_a&0&V_2&0\\
0&V_a&0&V_2\\
V_2&0&V_b&0\\
0&V_2&0&V_b
\end{array}\right)\left(\begin{array}{cccc}
0&0&0&0\\
\Pi^{aa}_{32}&\Pi^{aa}_{33}&0&0\\
0&0&0&0\\
0&0&\Pi^{bb}_{32}&\Pi^{bb}_{33}.
\end{array}\right).\nonumber\\
\eea
where $\mathbb{1}_4$ is $4\times 4$ identity matrix.

\subsubsection{Role of $V_{pp}$}
As we did in one-band case, we first present the form of $\chi_R(\Omega)$ neglecting $[V_{ph}]$ [the last term in Eq. (\ref{eq:21})].  We call this quantity $\chi_R^{pp}$. Using Eq. (\ref{eq:fan55}), we obtain
\bea\label{eq:res_a}
-\chi_R^{pp}(\Omega)&=&(c^a_1)^2\left[\Pi^{aa}_{33}+
\frac{(\Pi^{aa}_{23})^2\left\{V_a \left(-2 + V_b\Pi^{bb}_{22}\right) -\Pi^{bb}_{22}V_3^2\right\}}{4\mathcal{D}}\right]+ (c^b_1)^2\left[\Pi^{bb}_{33}+
\frac{(\Pi^{bb}_{23})^2\left\{V_b \left(-2 + V_a\Pi^{aa}_{22}\right)-\Pi^{aa}_{22}V_3^2\right\}}{4\mathcal{D}}\right]\nonumber\\
&&- 2(c^a_1)(c^b_1)
\left[\frac{\Pi^{aa}_{23}\Pi^{bb}_{23}V_3}{2\mathcal{D}}\right],
\eea
where, $\mathcal{D}=(1-V_a\Pi^{aa}_{22}/2)(1-V_b\Pi^{bb}_{22}/2)-\Pi^{aa}_{22}\Pi^{bb}_{22}V_3^2/4$. The polarization operators can be written in terms of $F_{a,b} (\Omega)$, $(\Omega/2\Delta^{a,b})^2$, and $l^{a,b}$. We express $l^a$ and $\l^b$ in terms of $V$ and $\Delta$ using the self-consistency condition $(1 + V_a l^a)(1+ V_b l^b) = V^2_3 l^al^b$ and the expression for the ratio of the gaps $\Delta^a/\Delta^b = - V_3 l^b/(1+ V_a l^a)$. After some algebra we obtain
\bea\label{eq:dsffs}
\chi^{pp}_{R}(\Omega)&=&-(c^a_1)^2F_a - (c_1^b)^2F_b + \frac{\kappa(c^a_1F_a+c^b_1 F_b)^2+\Omega^2\left\{(c^a_1)^2F_a^2F_b+(c_1^b)^2F_b^2F_a\right\}}{\kappa(F_a+F_b)+\Omega^2F_aF_b} \nonumber \\
&=&\frac{\kappa(c_1^a-c_1^b)^2}{\Omega^2+\kappa\left(\frac{1}{F_a}+\frac{1}{F_b}\right)},~~\text{where}~\kappa=\frac{8V_3\Delta^a\Delta^b}{V_aV_b-V_3^2}.
\eea
\end{widetext}
Eq. (\ref{eq:dsffs}) has been recently  obtained in Ref. \onlinecite{Lara} using a gauge-invariant  effective action formalism.
Note that the first two terms  $(c_1^a)^2F_a+(c_1^b)^2F_b$ account for the contribution from particle-hole bubbles without vertex corrections. Assuming that $\Delta^a<\Delta^b$, this portion of $\chi^{pp}_R (\Omega)$ has an edge singularity at $\Omega = 2\Delta^a +0$, because at  $\Omega=2\Delta^a(1+x)$, $F_a (\Omega)$ behaves as $F_a(\Omega)\approx 2 i\nu_{F,a}/\sqrt{2x}$.  When vertex corrections due to $[V_{pp}]$ are included, the edge singularity cancels, and the Raman intensity behaves as
\bea\label{eq:ff}
{\rm Im}[\chi^{pp}_R(\Omega)]&\approx&\frac{\kappa^2(c^a_1-c^b_1)^2}{2\nu_a([2\Delta^a]^2+\kappa/\bar F_b)^2}{\rm Re}[\sqrt{2x}],
\eea
where $\bar F_b$ denotes the real number $F_b(2\Delta^a)$. The removal of the edge singularity is illustrated in Fig. \ref{fig:ex}.

At $\Omega < 2 \Delta^a$  Raman intensity $R(\Omega) \propto {\rm Im} \chi^{pp}_R (\Omega)$  generally vanishes, but may have a  $\delta$-function peak
if the denominator in Eq. (\ref{eq:dsffs}) has a pole at some frequency from this range.   The pole position is determined from $\Omega^2+\kappa(1/F_a+1/F_b)=0$. The corresponding collective mode is the Leggett mode.\cite{Leggett,Lara,einzel} We see that this mode is indeed Raman active. Because $F_{a,b}>0$, the mode exists when $\kappa<0$ and $4(\Delta^a)^2 > |\kappa|/F_b (\Omega =2\Delta^a)$. The condition $\kappa <0$ implies that SC  is driven by intra-band pairing, i.e. $V_a V_b> V_3^2$, and, simultaneously, $V_a, V_b <0$ (otherwise there would be no attraction). For interband-driven SC, when $V_aV_b<V_3^2$, $\kappa>0$, and there is no Leggett mode, hence no $\delta-$function peak in $\chi_R^{pp} (\Omega)$ at $\Omega < 2\Delta^a$.

When $c_1^a=c_1^b$, $\chi^{pp}_R (\Omega)$ vanishes. This is again the consequence of particle number conservation, like in the one-band case. The  $(c_1^a-c_1^b)^2$ factor in $R(\Omega)$ has been obtained in Ref.~\onlinecite{Girsh}, but  attributed to  partial screening of the Coulomb interaction in a two-band SC.  We have shown that this factor appears due to vertex corrections in the particle-particle channel even before renormalization by the Coulomb interaction is considered.

\subsubsection{Role of $V_{ph}$}
We now include $V_{ph}$. The full expression for $\chi_R (\Omega)$ obtained from Eqs. (\ref{eq:fan55}) and (\ref{eq:21}) is rather long. To keep the formulas short, we assume $V_a=V_b=V_4$ and set $\nu_{F,a} = \nu_{F,b} = \nu_F$.  This leads to $\Delta^a=-s \Delta^b = \Delta$, where $s = {\rm sgn}(V_3)$. Then $\Pi_{22,33}^{aa}=\Pi_{22,33}^{bb}=\Pi_{22,33}$ and $\Pi_{23}^a=-s\Pi_{23}^b=\Pi_{23}$. With these simplifications, the Raman susceptibility is  given by
\beq\label{eq:res_1}
\chi_{R} (\Omega)= \frac{(c_1^a-c_1^b)^2}{2}\left[\frac{\Pi_{33}-\frac{(\Pi_{23})^2}{\frac{2}{V_4+ |V_3|}-\Pi_{22}}}
{1+\frac{V_{4}-sV_2}{2}\left(\Pi_{33}-\frac{(\Pi_{23})^2}{\frac{2}{V_4+|V_3|}-\Pi_{22}}\right)}\right].
\eeq
Substituting the expressions for $\Pi_{ij}$ in terms of $F(\Omega)$ using Eq. (\ref{eq:ds}), we obtain
\beq\label{eq:ll}
\chi_R (\Omega)=\frac{(c_1^a-c_1^b)^2\kappa F(\Omega)}{(\Omega^2-(V_4-sV_2)\kappa)F(\Omega)+2\kappa}.
\eeq
Comparing this formula to the one for $\chi^{pp}_R(\Omega)$ [Eq. (\ref{eq:dsffs})], we see that (i) vertex corrections in the particle-hole channel shift the position of the Leggett mode, when this mode exists, and (ii) may also give rise to another, excitonic-like collective mode, when $(V_{4}-sV_2) F(\Omega) \approx 2$ At the frequency of this collective mode, Raman intensity $R(\Omega) \propto \text{Im} [\chi_R (\Omega)]$ has a $\delta-$function peak. The interplay between the Leggett mode due to vertex corrections in the particle-particle channel and the excitonic mode due to vertex corrections in the particle-hole channel  requires a more detailed analysis of the structure of the denominator in Eq. (\ref{eq:res_1}). This is what we do next.

\begin{figure}[htp]
$\begin{array}{cc}
\includegraphics[width=0.5\columnwidth]{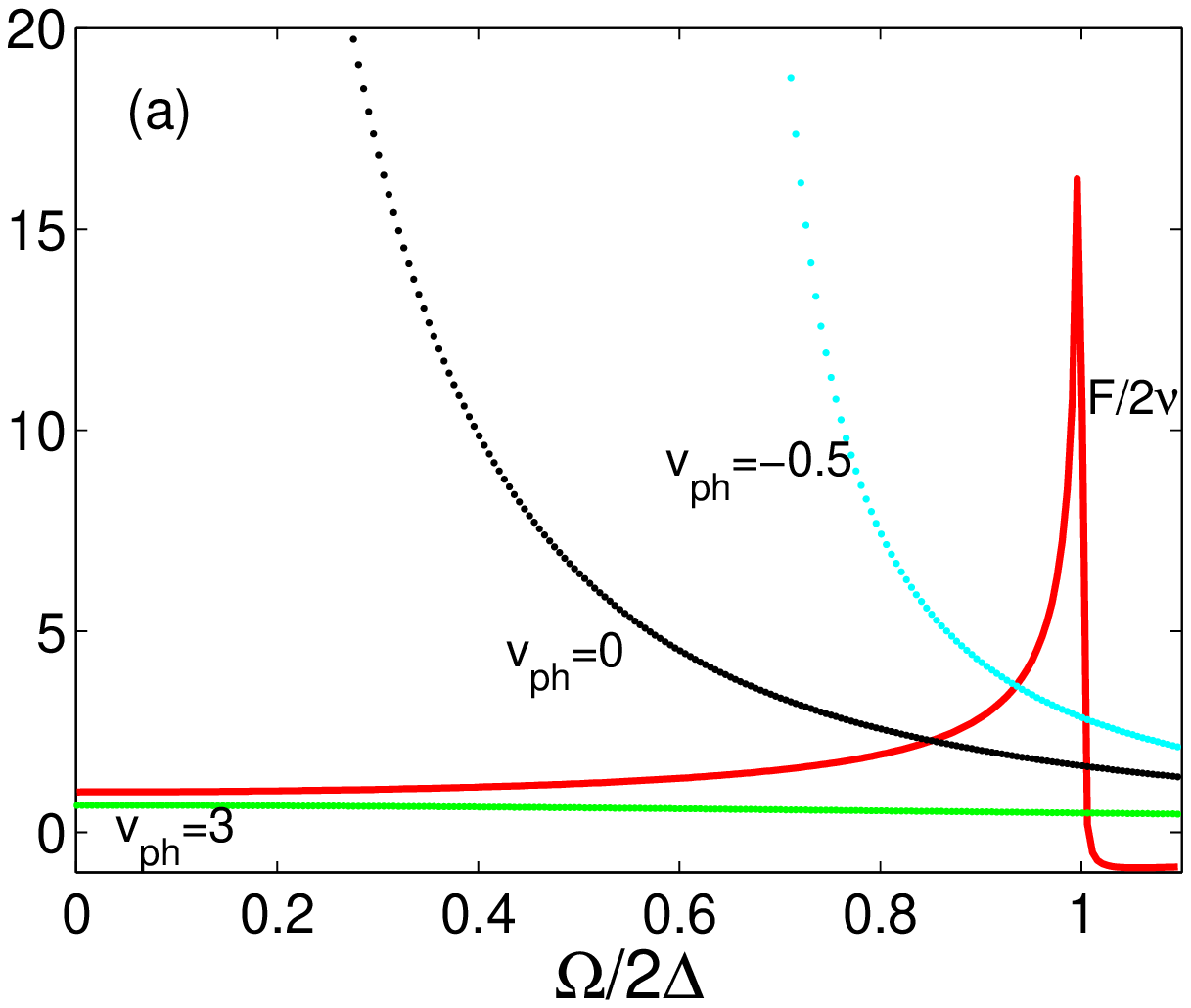}&
\includegraphics[width=0.5\columnwidth]{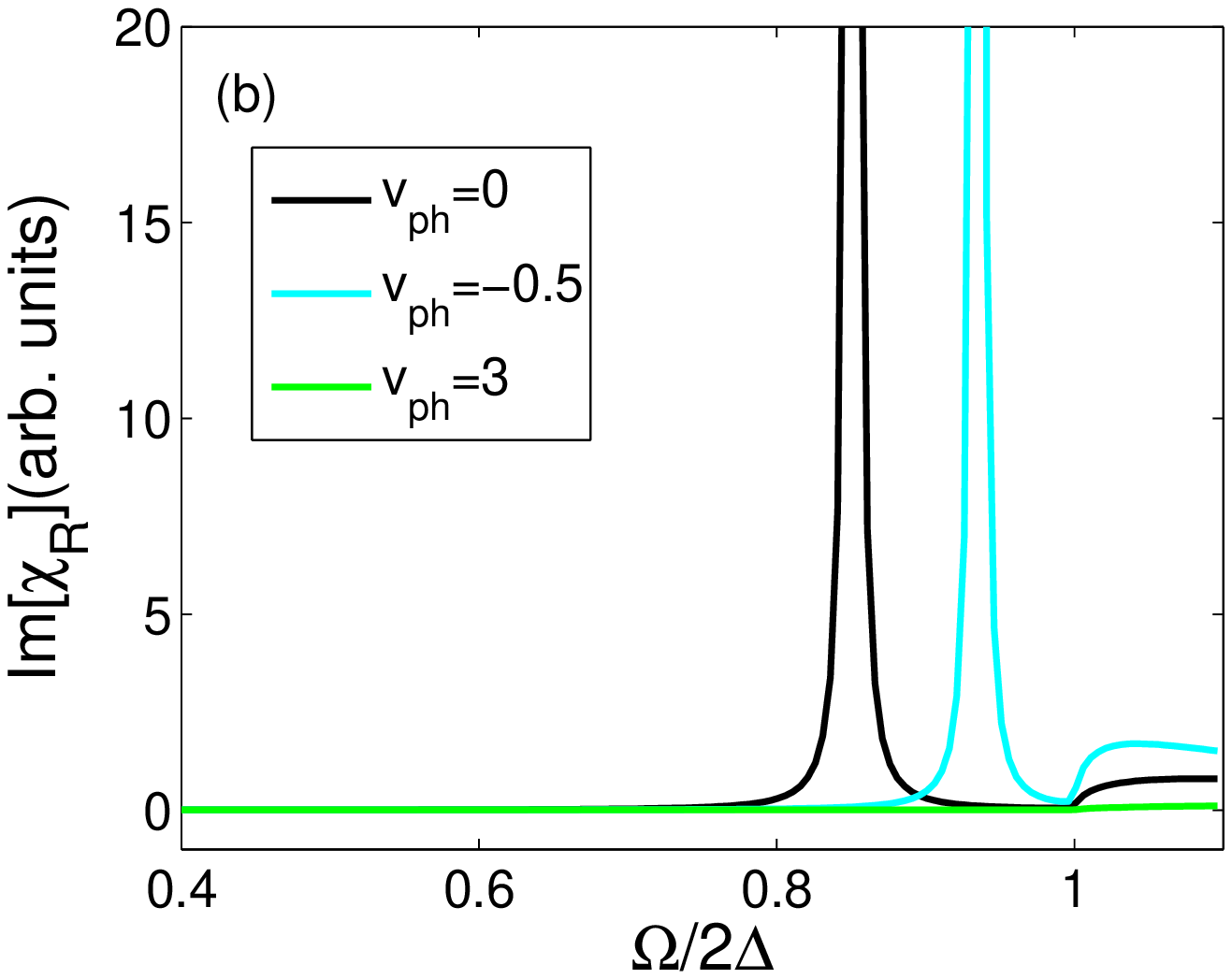}\\
\includegraphics[width=0.5\columnwidth]{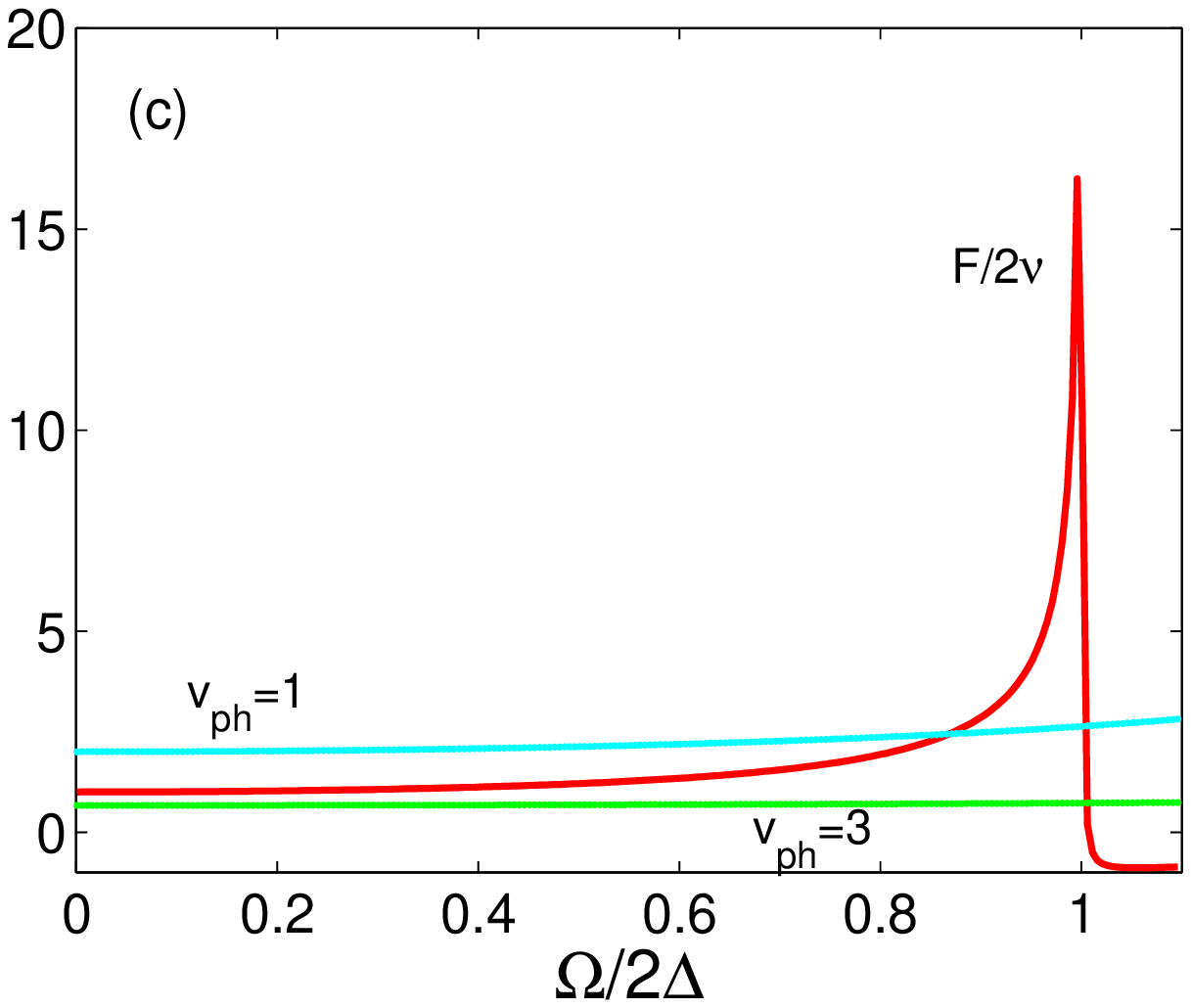}&
\includegraphics[width=0.5\columnwidth]{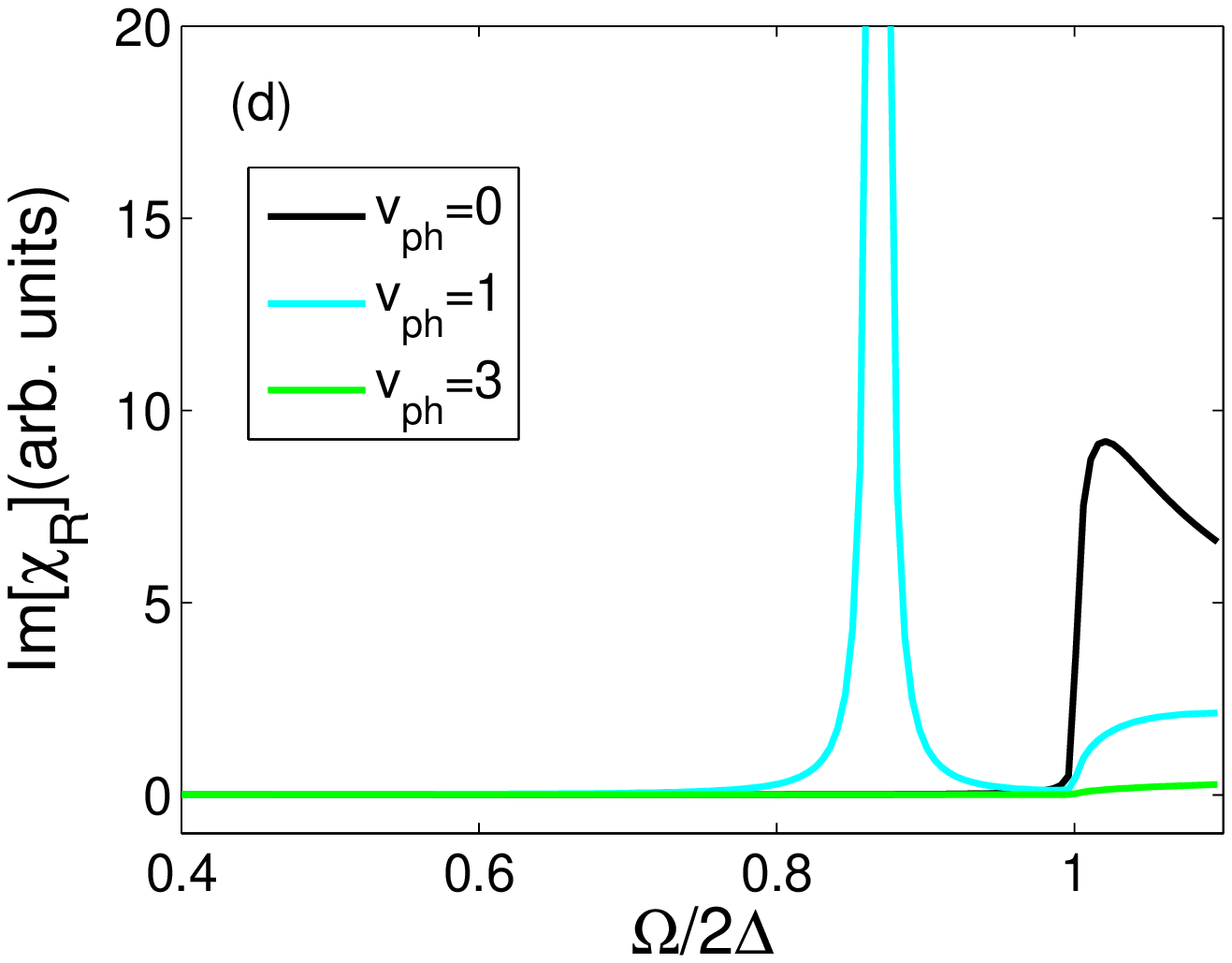}
\end{array}$
\caption{
\label{fig:33} Color online: (a) Solutions to the transcendental Eq. (\ref{eq:ll2}) given by the intersection of $\tilde F = F/2\nu$ with RHS of Eq. (\ref{eq:ll2}) for different $v_{ph}$. (b) The Raman spectrum for $v_{ph}=0,~v_{ph}=-0.5\in \{-1/|\tilde\kappa|,2\}$, and $v_{ph}=3$ (not in that bound). This case corresponds to the renormalization of the Leggett mode. Here $\tilde\kappa=-0.83$. (c) Solutions to the transcendental Eq. \ref{eq:ll2} and (d) the Raman spectrum for $v_{ph}=0,~v_{ph}=1\in \{1/\tilde\kappa,2\}$, and $v_{ph}=3$ (not in that bound). This shows the emergence of a possible excitonic mode. Here $\tilde\kappa=4.2$. Note the removal of the $2\Delta$ edge singularity in all cases.}
\end{figure}

\subsubsection{Interplay between the Leggett mode and the excitonic mode}
To be specific, define  $V_{pp}=V_4+|V_3|$, $V_{ph}=V_4-sV_2$, and $V_{sc} = V_4 - |V_3|$. We keep $|\Delta^a| = |\Delta^b|= \Delta$. For  $V_{ph}=0$, the  Leggett mode necessarily exists at $\Omega < 2\Delta$ as long as $V_{pp} < 0$ and $V_{sc} <0$ (intra-band attraction-driven superconductivity) and is located at a frequency at which $\Omega^2F(\Omega) =2|\kappa|$. In the presence of $V_{ph}$ the zero of the denominator in Eq. (\ref{eq:res_1}) shifts to
\beq
(\Omega^2-V_{ph}\kappa)F(\Omega) = -2\kappa
\eeq
It is convenient to re-express this relation in dimensionless variables $\tilde F(\Omega)\equiv F(\Omega)/2\nu_F$, $\tilde{\Omega}\equiv \Omega/2\Delta$,  $\tilde\kappa = \kappa/(4\Delta^2 \nu_F)$, and $v_{pp} = V_{pp} \nu_F$, etc.,  and $\tilde\kappa   =  -(v_{pp}-v_{sc}/(v^2_{pp} -v^2_{sc})$. The locations of the poles are then the roots of the transcendental equation
\beq\label{eq:ll2}
\tilde F(\Omega)=\frac{2\tilde \kappa}{-\tilde\Omega^2 + v_{ph}\tilde\kappa}.
\eeq
A straightforward analysis shows that when $-1/|\tilde \kappa|<v_{ph}<2$, there is  one solution at $\tilde \Omega <1$: the renormalized Leggett mode. If $v_{ph}$ is outside those bounds, there is no solution on the real frequency axis  (see Fig. \ref{fig:33}a and b). Now, if $\tilde{\kappa}>0$ and  min$\{1/\tilde\kappa,2\}<v_{ph}<$max$\{1/\tilde\kappa,2\}$, there is again one solution at $\tilde \Omega <1$. This is an excitonic mode (see Fig. \ref{fig:33}c and d). When $v_{ph}$ is outside those bounds, there is again no solution on the real frequency axis.
\begin{figure*}[htp]
$\begin{array}{ccc}
\includegraphics[width=0.65\columnwidth]{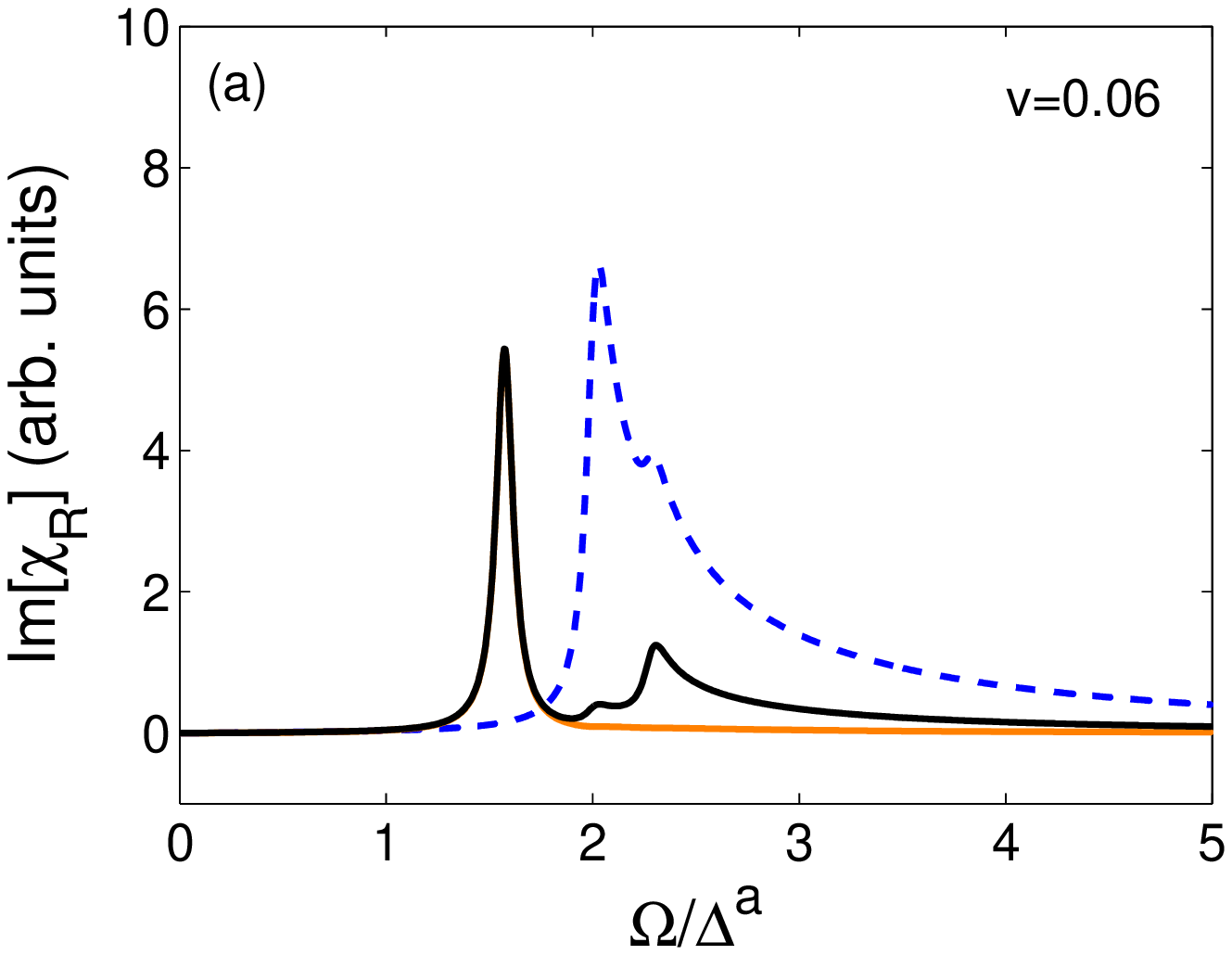}&
\includegraphics[width=0.65\columnwidth]{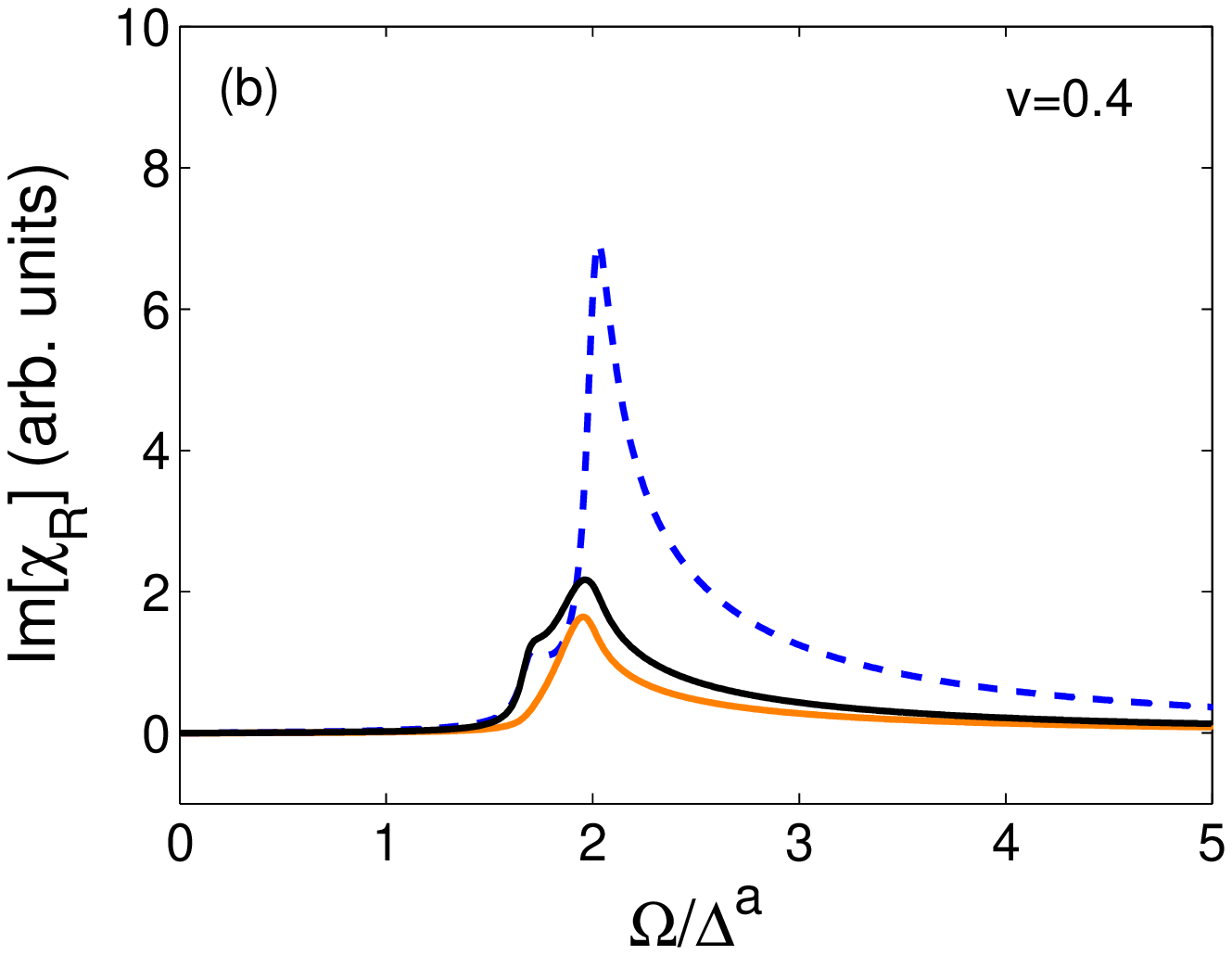}&
\includegraphics[width=0.65\columnwidth]{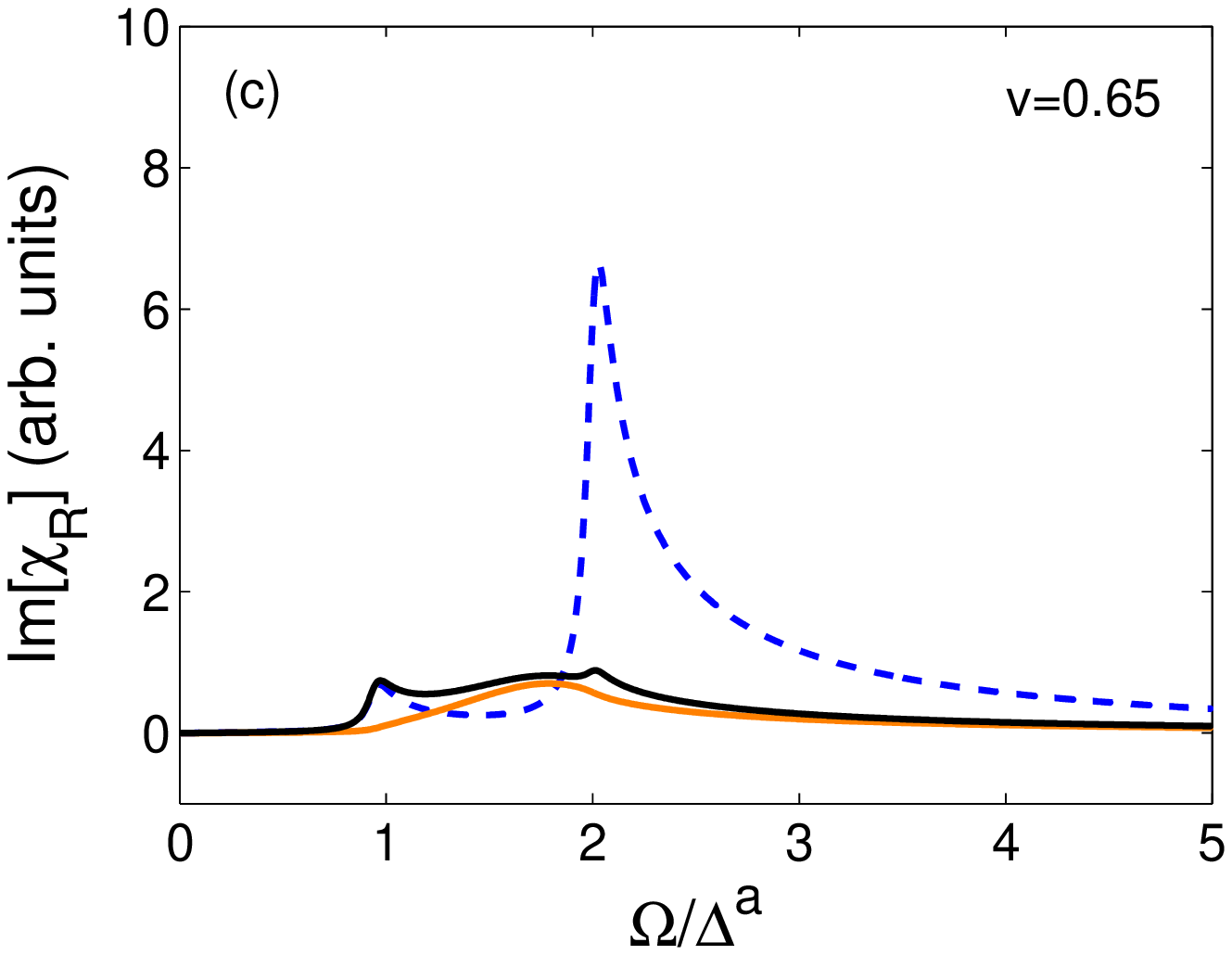}
\end{array}$
\caption{
\label{fig:22} The Raman response in the A$_{1g}$ channel for a 2-band $s-$wave SC. The dashed line is the contribution from the bare bubble. The light red line is the response after accounting for vertex corrections with constant $\gamma_{\bk}$: $c^a_1=1$, $c^b_1=0.3$. The dark line shows the Raman response for a non-constant $\gamma_{\bk}$ where we added $c^a_2=0.2$, $c^b_2=0.3$ (corresponding to the $\cos4\theta$ harmonic). (a) and (b) refer to the case of intra-band driven attractive SC ($V_aV_b-V_3^2>0$) and show how the Leggett mode can be pushed to the continuum if $V_3$ is increased. (c) is the case of inter-band driven attractive SC ($V_aV_b-V_3^2<0$) which has no Leggett mode. Note the removal of $2\Delta$ edge singularity and strong suppression of the spectral weight around $2\Delta$ in all cases. Here $v\equiv V_3\nu_b$; $\nu_a/\nu_b=0.6$; and $V_a\nu_b=-0.5$, $V_b\nu_b=-0.35$ for (a) and (b); $V_a\nu_b=0.3$, $V_b\nu_b=0.6$ for (c). A fermionic lifetime of $0.05\Delta$ is added for broadening.}
\end{figure*}

\subsection{Case of anisotropic two-band system}\label{sec:a2bandex}
We now include one more harmonic into $\gamma^i(\bk)$ for each band: $\gamma^a_{\bk} = c^a_1 f^1_{\bk} + c^a_2 f^2_{\bk}$, $\gamma^b_{\bk} = c^b_1 f^1_{\bk} + c^b_2 f^2_{\bk}$, and include
$V^{aa,22}_{i}$ and $V^{ab,22}_{i}$ harmonics into the interaction.  For brevity, we  denote $V^{aa,22}_{4} = V^{bb,22}_{4} = {\tilde V}_4$, $V^{ab,22}_{2} = {\tilde V}_2$, and $V^{ab,22}_{3}= {\tilde V}_3$. The $[V]$, $[\Pi]$ and $[c]$ matrices in this case are:
\begin{widetext}
\beq\label{eq:fan55A}
[V_{pp}]=\left(
\begin{array}{cccc}
V_4&V_3&0&0\\
V_3&V_4&0&0\\
0&0&\tilde V_4&\tilde V_3\\
0&0&\tilde V_3&\tilde V_4\\
\end{array}
\right)\otimes\sigma_0;~
[V_{ph}]=\left(
\begin{array}{cccc}
V_4&V_2&0&0\\
V_2&V_4&0&0\\
0&0&\tilde V_4&\tilde V_2\\
0&0&\tilde V_2&\tilde V_4\\
\end{array}
\right)\otimes\sigma_0;
\eeq
and $[c]^T=(0,c_1^a,0,c_1^b,0,c_2^a,0,c_2^b)$. Following usual steps we obtain,
\bea\label{eq:fan337}
\chi_R (\Omega) &=&
\frac{(c_1^a-c_1^b)^2}{2}\left[\frac{\Pi_{33}-\frac{(\Pi_{23})^2}{\frac{2}{V_4+ |V_3|}-\Pi_{22}}}
{1+\frac{V_{4}-sV_2}{2}\left(\Pi_{33}-\frac{(\Pi_{23})^2}{\frac{2}{V_4+|V_3|}-\Pi_{22}}\right)}\right] \nonumber \\
&& +
\frac{(c_2^a+c_2^b)^2}{2}\left[\frac{\Pi_{33}-\frac{(\Pi_{23})^2}{\frac{2}{\tilde V_4-s\tilde V_3}-\Pi_{22}}}
{1+\frac{\tilde V_{4}+s\tilde V_2}{2}\left(\Pi_{33}-\frac{(\Pi_{23})^2}{\frac{2}{\tilde V_4-s\tilde V_3}-\Pi_{22}}\right)}\right]+\frac{(c_2^a-c_2^b)^2}{2}\left[\frac{\Pi_{33}-\frac{(\Pi_{23})^2}{\frac{2}{\tilde V_4+s\tilde V_3}-\Pi_{22}}}
{1+\frac{\tilde V_{4}-s\tilde V_2}{2}\left(\Pi_{33}-\frac{(\Pi_{23})^2}{\frac{2}{\tilde V_4+s\tilde V_3}-\Pi_{22}}\right)}\right].\nonumber\\
\eea
\end{widetext}
The particle conservation does not impose  restrictions  in the $c_2$ channel, hence both terms in the second line in Eq. (\ref{eq:fan337}) are generally  non-zero. There may be additional resonances in $\chi_R(\Omega)$ due to poles in these two terms.  It is essential to note that each term in Eq. (\ref{eq:fan337}) still vanishes at $\Omega =2\Delta$, i.e., the divergence in Im[$\Pi_{33}$] at $\Omega = 2\Delta+0$ is eliminated by vertex corrections. In practice, higher harmonics are more likely to just add the spectral weight to the $2\Delta$ Raman continuum rather than induce new resonances below $2\Delta$  (see Fig. \ref{fig:22}).

\section{Role of Coulomb interaction}\label{sec:screening}
We now move to include the effects  of Coulomb interaction into the A$_{\rm 1g}$ response. It was argued in the past that screening from the long-range Coulomb interaction is a necessary ingredient of Raman analysis, and that this screening accounts for the vanishing of the Raman susceptibility in the case when Raman vertex $\gamma^a (\bk)$ can be treated as a constant, independent of the band index. We have now shown that vertex corrections already account for the vanishing of  $\chi_R (\Omega)$ in this situation. We now argue that, at $\bq \to 0$, the Coulomb interaction does not affect the Raman susceptibility for an arbitrary Raman vertex $\gamma^a (\bk)$. Namely, we argue that both $\pi_{CR} (= \pi_{RC})$ and $\pi_{CC}$ vanish, no matter what $\gamma^a (\bk)$ is and thus there is no screening correction to the Raman response in a SC when $q\rightarrow 0$.

For non-A$_{\rm 1g}$ scattering geometry, the vanishing of  $ \pi_{RC,CR}$ is obvious because non-s-wave eigenfunctions and a constant charge form-factor are orthogonal. For s-wave scattering and $\gamma^a (\bk)$ = const it is also obvious because then $\pi_{RC,CR} = \pi_{RR} = \pi_{CC}$, and all vanish for $q \to 0$.  However, it is less obvious when the scattering is in s-wave geometry (e.g., $A_{1g}$ geometry for a 2D quare lattice), but $\gamma^a (\bk)$ has momentum-dependent harmonics, or has momentum-independent, but different values, for different bands.

The logical reasoning for the vanishing of  $ \pi_{RC,CR}$ in this case is the following. The screening correction is given by $\pi_{RC}\pi_{CR}/[1/V_C(\bq)-\pi_{CC}]$. The quantity $\pi_{CC}$ is a fully renormalized density-density correlator, and it vanishes at $q=0$ (as we just demonstrated for the Raman bubble in case when $\gamma^a (\bk) $ = const. One can easily extend that analysis to finite $q$ and show that at $v_Fq\ll\Omega$, $\pi_{CC}\propto q^2(\Omega_{\rm pl}/\Omega)^2$, where $\Omega_{\rm pl}$ is the plasma frequency. This holds both in the normal state and in the SC state.  The obvious consequence is that $1/V_C (\bq) - \pi_{CC} =0$  at the plasma frequency in 3D and at $\Omega \propto \sqrt{q}$ in 2D.  Because in both cases $1/V_C (\bq) - \pi_{CC}$ vanishes at $q=0$,  the screening correction would have an unphysical divergent contribution to the Raman response if  $\pi_{RC,CR}$ were non-zero at $q=0$.  The requirement that the theory must be free from divergencies then forces $\pi_{RC,CR} (q=0)$ to vanish. The expansion of a charge response function in $q$ in necessarily analytic,\cite{maslov} hence $\pi_{RC,CR}(q) \propto q^2$.  Then the contribution to $R(\Omega)$ from Coulomb screening scales as $q^2$ in 3D and as $q^3$ in 2D and vanishes at $q \to 0$.

Below we demonstrate that $\pi_{RC,CR} (q=0) =0$ for the two non-trivial cases -- a two-band superconductor with different momentum-independent $\gamma^a$ for the two bands and a one-band superconductor with an anisotropic s-wave gap and arbitrary $\gamma (\bk)$ for s-wave scattering.

\subsection{Absence of screening in a two-band SC with isotropic gap}
The generic expression for $\pi_{RC}(Q)$ is given by Eq. (\ref{eq:Raman2}), where, we remind the reader, $\bar \Gamma^a$ is the fully renormalized charge vertex, with partial components ${\bar \Gamma}^{a,t}_2$  and  ${\bar \Gamma}^{a,t}_3$. Because we assume the Raman vertex to be momentum-independent, only partial components with $t=1$ are non-zero.

To show that $\pi_{RC} (q=0)$ vanishes, we follow the same strategy as for the analysis of $\pi_{RR}$ and first include only the renormalizations in the particle-particle channel, i.e., neglect terms with $\Pi^{aa,11}_{32}, \Pi^{aa,11}_{33}, \Pi^{bb,11}_{32}$ and $\Pi^{bb,11}_{33}$ (we call the corresponding piece $\pi^{pp}_{RC}$.
From Eq. (\ref{eq:Raman2}) we then obtain
\bea\label{eq:idk3}
\pi^{pp}_{RC} (q=0) &=&c^a_1\left[\Pi^{aa,11}_{33}\bar\Gamma^{a,1}_{3} + \Pi^{aa,11}_{32}\bar\Gamma^{a,1}_{2}\right]\nonumber\\
&&+c^b_1\left[\Pi^{bb,11}_{33} \bar\Gamma^{b,1}_{2} + \Pi^{bb,11}_{32}\bar\Gamma^{b,1}_{2}\right]
\eea
Evaluating $\Pi^{aa,11}_{32}(\Omega)$ and $\Pi^{aa,11}_{33}(\Omega)$, we find that they are related:
\beq\label{eq:wtf}
\Pi^{aa,11}_{32}(\Omega)=\frac{i\Omega}{2\Delta^a}\Pi^{aa,11}_{33}(\Omega)
\eeq
The same relation holds for the $b-$band
\beq
\Pi^{bb,11}_{32}(\Omega)=\frac{i\Omega}{2\Delta^b}\Pi^{bb,11}_{33}(\Omega)
\eeq
Substituting these relations into Eq. (\ref{eq:idk3}), we obtain
\bea\label{eq:idk13}
\pi^{pp}_{RC} (q=0) &=&c^a_1\Pi^{aa,11}_{33}\left[\bar\Gamma^{a,1}_{3}+ (i\Omega/2\Delta^a)\bar\Gamma^{a,1}_{2}\right]\nonumber\\
&&+c^b_1\Pi^{bb,11}_{33}\left[\bar\Gamma^{b,1}_{3} + (i\Omega/2\Delta^b)\bar\Gamma^{b,1}_{2}\right]
\eea
Now, from the last two lines in Eq. (\ref{ac_2}) and Eq. (\ref{eq:eq}) we obtain for ${\bar \Gamma}$:
\bea\label{eq:last}
\bar\Gamma^{a,1}_3&=&1 \nonumber \\
\bar\Gamma^{a}_{2,1}&=&\frac{2V_3\Pi^{bb,11}_{23}-\Pi^{aa,11}_{23}\left\{V_a \left(-2 + V_b\Pi^{bb,11}_{22}\right) -\Pi^{bb,11}_{22}V_3^2\right\}}{4\mathcal{D}}\nonumber\\
&=&-\frac{2\Delta^a}{i\Omega},\\
\bar\Gamma^{b,1}_3&=&1 \nonumber \\
\bar\Gamma^{b,1}_{2}&=&\frac{2V_3\Pi^{aa,11}_{23}-\Pi^{bb,11}_{23}\left\{V_b \left(-2 + V_a\Pi^{aa}_{22}\right) -\Pi^{aa,11}_{22}V_3^2\right\}}{4\mathcal{D}}\nonumber\\
&=&-\frac{2\Delta^b}{i\Omega} \nonumber
\eea
Substituting into Eq. (\ref{eq:idk13}) we see that {\it each term} in Eq. (\ref{eq:idk13}) vanishes.  Then $\pi^{pp}_{RC} (q=0) =0$ for arbitrary $c^a_1$ and $c^b_1$.

The analysis can be straightforwardly extended to include the renormalizations in the particle-hole channel. We indeed found that the result holds, i.e., $\pi_{RC} (q=0) =0$. We do not show the details of the proof as the calculations are somewhat lengthy.

\subsection{Absence of screening in a one-band SC with anisotropic gap and arbitrary $\gamma (\bk)$}

To be specific, consider a 2D SC on a square lattice and assume that the interaction is only in the pairing channel and is in the form $V_{pp}(\bk,\bk')=Vf_{\bk} f_{\bk'}$, where $f_{\bk} = 1 + r\cos4\theta+...$. We assign the index $f$ to this harmonic (i.e., set $f_{\bk} \equiv f^f_{\bk}$) and the index 1 to $f^1_{\bk} =1$. For such $V_{pp}$, the pairing gap has the form $\Delta (\bk) = \Delta_0f_{\bk}$. Using Eq. (\ref{ac_2}), one can easily check that in this situation $\bar\Gamma_3 =1$ because, the renormalization of ${\bar \Gamma}_3$ could only come from the interaction in the particle-hole channel. Further,  $\bar\Gamma_2(\bk)=\sum_t \bar\Gamma^t_2(\bk) f^t_{\bk} $ has only the harmonic with the index f, i.e.,  $\bar\Gamma_2(\bk) = {\bar\Gamma}_2 f_{\bk}$. Substituting the last form into Eq. (\ref{ac_2}) we obtain, skipping the band index,
\bea\label{eq:idk4}
\bar\Gamma_2&=&\frac{V}{2}\Pi^{ff}_{22}\bar\Gamma_2 +\frac{V}{2}\Pi^{f 1}_{23}
\eea
Hence
\bea\label{eq:qwerty}
\bar\Gamma_2&=&\frac{\Pi^{f 1}_{23}}{\frac2V-\Pi^{ff}_{22}}.
\eea
From Eq.  (\ref{eq:wtfm}) we then obtain, for arbitrary $\gamma (\bk) = \sum_t c_t f^t_{\bk}$
\bea\label{eq:mmm}
\pi_{RC}(\Omega)&=&\sum_t c_t \int_{K'}f^t_{\bk'} f_{\bk'}\text{Tr}[\sigma_3 G_{K'}\sigma_2G_{K'+Q}]\bar\Gamma_2\nonumber\\
&&+ \sum_t c_t \int_{K'}f^t_{\bk'}\text{Tr}[\sigma_3 G_{K'}\sigma_3G_{K'+Q}] \nonumber\\
&\equiv& \sum_t c_t \left(\Pi^{t f}_{32}\bar\Gamma_2 + \Pi^{ t 1}_{33}\right).
\eea
On explicitly evaluating the polarization operators $\Pi^{ff}_{22}, \Pi^{f 1}_{23}, \Pi^{t f}_{32}$, and $\Pi^{ t 1}_{33}$, we obtain
\bea\label{eq:idk5}
\Pi^{f 1}_{23}&=&\frac{i\Omega}{2\Delta_0}F^{f f}(\Omega);\nonumber\\
\Pi^{t f}_{32}&=&-\frac{i\Omega}{2\Delta_0} \tilde F^{t f}(\Omega);\nonumber\\
\Pi^{t 1}_{33}&=&-\tilde F^{t f}(\Omega);\nonumber\\
\Pi^{f f}_{22}&=&\frac2V-\left(\frac{\Omega}{2\Delta_0}\right)^2 F^{f f}(\Omega),
\eea
where $F^{ff}(\Omega)\equiv\int_{\bk}\Delta_0^2f^2_{\bk}/4E[E^2-(\Omega/2)^2]$ and $\tilde F^{tf}(\Omega)\equiv\int_{\bk}\Delta_0^2 f^t f^2_{\bk}/4E[E^2-(\Omega/2)^2]$.
Observe that
\beq\label{eq:sm2}
\Pi^{t f}_{32}=\frac{i\Omega}{2\Delta_0}\Pi^{t 1}_{33}.
\eeq
Using the last relation we re-express  $\pi_{RC}$ from Eq. (\ref{eq:mmm}) as
\beq\label{eq:sm3}
\pi_{RC}(\Omega)=\sum_t c^t \Pi^{t 1}_{33}\left[1+\frac{i\Omega}{2\Delta_0}\bar\Gamma_2\right].
\eeq
 Using Eqs. (\ref{eq:qwerty}) and (\ref{eq:idk5}) we then find that
\beq\label{eq:sm1}
\bar\Gamma_2= - \frac{2\Delta_0}{i\Omega}.
\eeq
Substituting this result in Eq. (\ref{eq:sm3}) we see each term under the sum over $t$ vanishes. As the consequence, $\pi_{RC}(\Omega)=0$ for \emph{any} $\gamma_{\bk}$.


\section{Non-A$_{\rm 1g}$ channels}\label{sec:learn}
Finally we briefly discuss the spectrum in non-A$_{\rm 1g}$ channels and highlight their appealing aspects for the experiments. To get there, let us note that the common features of the Raman response across all A$_{\rm 1g}$ and non-A$_{\rm 1g}$ channels are: (1) removal of the $2\Delta$ edge singularity and (2) presence of collective modes under favorable conditions. What is different is that in A$_{\rm 1g}$ scattering geometry one always probes  fluctuations in the pairing channel with the symmetry of the pairing gap, whereas in the non-A$_{\rm 1g}$ scattering geometry one probe fluctuations in  subleading channels, where interaction may be attractive (but weaker than in the leading channel). For example,  in an $s-$wave SC  Raman intensity in A$_{\rm 1g}$ scattering geometry may have sharp peaks corresponding to Leggett modes, which represent fluctuations of the relative phases of the multi-component order parameters,  while  Raman intensity in B$_{\rm 1g}$ scattering geometry may have peaks corresponding to BS modes if a subleading $d-$wave channel is attractive.

It is also easy to see that in a $d$-wave SC, it is the Raman scattering in A$_{\rm 1g}$ geometry that is strongly affected by vertex corrections. Indeed, to convert from a particle-hole to a particle-particle channel, one needs a combination of a normal and an anomalous Green's function. The anomalous Green's function has $\Delta (\bk)$ in the numerator. Hence, for A$_{\rm 1g}$ scattering geometry, the resulting form-factor for particle-particle channel has a $d-$wave symmetry. Then one needs $d-$wave pairing interaction (the same that leads to SC) to renormalize it.  In other words, the fluctuations of a $d-$wave SC order parameter  are probed in Raman experiments in an A$_{\rm 1g}$ scattering geometry. And vise versa - the Raman response in a B$_{\rm 1g}$ scattering geometry in a $d$-wave SC probes pairing fluctuations in $s-$wave channel.

\section{Conclusion}
\label{sec:conclusions}

To conclude, we have presented the general scheme to calculate the
Raman response in a multi-band superconductor with both
short-range and long-range interactions between fermions. We
grouped the interactions into the screening part and the part
which accounts for the renormalizations of the Raman vertex.  We
further decomposed vertex renormalizations into those in the
particle-particle  and in the particle-hole channels. The
renormalizations in the particle-particle channel are unavoidable
in a superconductor because a particle-hole Raman vertex can be
converted into particle-particle vertex via the renormalization
which involves one normal and one anomalous fermionic Green's
function ($\Pi_{23}$ polarization bubble in the Nambu formalism).
We have presented a general  formula that accounts for vertex
corrections in particle-particle and particle-hole channels in a
generic multi-band superconductor. We demonstrated  that vertex
corrections in the  particle-particle channel cannot be neglected
as they enforce the constraint imposed by the particle number
conservation. This point has been emphasized in Ref.
[\onlinecite{Lara}], and our results are in agreement with theirs.

We argued that in a situation when the Raman form-factor
($\gamma_{\bk}$) is momentum-independent and the same for all
bands,  vertex corrections completely eliminate the Raman
response. For a generic polarizations of light  Raman  intensity
remains finite, but vertex corrections remove the $2\Delta$ edge
singularity of the bare bubble, leaving only a broad maximum at a
frequency \emph{above} $2\Delta$. Besides,  vertex corrections
account for $\delta-$function contributions to the Raman intensity
from collective modes, at frequencies below twice the minimum gap.
Specifically, we analyzed the contributions to $A_{1g}$  Raman
intensity of a 2D system on a square lattice from the Leggett mode
in the particle-particle channel and the excitonic mode in the
particle-hole channel, and the interplay between the two.

We also demonstrated that, once  vertex corrections inside the
Raman bubble are included, the remaining  RPA-type
renormalizations  of the Raman susceptibility by long-range
Coulomb interaction (screening corrections) are negligibly small
at $q \to 0$,  even for the A$_{\rm 1g}$ scattering geometry on a
lattice when $s-$wave Raman form-factor has some momentum
dependence.  This implies that long-range Coulomb interaction  is
irrelevant for the Raman scattering when $q\rightarrow 0$.

The formalism that we developed applies to any pairing symmetry,
any number of bands, and a generic form of the Raman vertex
$\gamma(\bk)$. It can also be  easily extended to tackle
time-reversal symmetry broken superconducting states, like
$s+is$,\cite{SM_AVC,Benfatto,spisLegget1,spisLegett2} $s
+id$,\cite{Hanke} etc.. The temperature dependence of the energies
of the collective modes can also be inferred  from Raman
experiments. To obtain it theoretically, one needs to keep
temperature dependence in the polarization operators
$\Pi^{aa,mn}_{ij}$.

\emph{Acknowledgements.}  We  thank G. Blumberg, L. Benfatto, T.
Devereaux, D. Einzel, R. Hackl, and D. Maslov  for useful
discussions. We are also thankful to L. Benfatto and R. Hackl  and
R. Hackl for comments on the manuscript. This work was supported
by the Office of Basic Energy Sciences, U.S. Department of Energy,
under awards DE-SC0014402 (AVC) and DE-FG02-05ER46236 (PJH).  AVC
thanks the Perimeter Institute for Theoretical Physics (Waterloo,
Canada) for hospitality during the final stages of this project.
The research at the Perimeter Institute is supported in part by
the Government of Canada through the Department of Innovation,
Science and Economic Development and by the Province of Ontario
through the Ministry of Research and Innovation. \vskip .5cm

\bibliographystyle{apsrev4-1}

\begin{thebibliography}{10}
\bibitem{note} We only retain intraband $d-$wave interaction $V^d$ to reduce the number of parameters.
\end{thebibliography}

\end{document}